\documentclass{ws-jca-noheader}

\usepackage[numbers,sort&compress]{natbib}
\usepackage{amssymb}
\usepackage{amsmath}
\usepackage{graphicx}
\usepackage{array}
\usepackage{ctable}
\usepackage{comment}
\usepackage{epstopdf}
\usepackage{cancel}
\usepackage{hyperref}

\newcommand{\DDelta}{\mathrm{\Delta}}
\newcommand{\FT}[1]{\mathcal{F}\left\{ #1 \right\}}
\newcommand{\IFT}[1]{\mathcal{F}^{-1}\left\{ #1 \right\}}
\renewcommand{\vec}[1]{\mathbf{#1}}

\begin{document}

\markboth{J. Jaros, A. Rendell \& B. Treeby}
{Full-wave nonlinear ultrasound simulation on distributed clusters}
\catchline{}{}{}{}{}

\title{Full-wave nonlinear ultrasound simulation on distributed clusters with applications in high-intensity focused ultrasound}

\author{Jiri Jaros\footnotemark[1]{ }\addtocounter{footnote}{1}\footnote{Previously at the Research School of Computer Science, The Australian National University, Canberra} , Alistair P. Rendell\footnotemark[3]{ }, and Bradley E. Treeby\footnotemark[4]{ }\addtocounter{footnote}{2}\footnote{Previously at the Research School of Engineering, The Australian National University, Canberra} \vspace{1em}}

\address{\footnotemark[1]{} Faculty of Information Technology, Brno University of Technology, \\ Bo\v{z}et\v{e}chova 2, 612 66 Brno, Czech Republic (jarosjir@fit.vutbr.cz)}
\address{\footnotemark[3]{} Research School of Computer Science, The Australian National University, \\ Canberra, ACT 0200, Australia (alistair.rendell@anu.edu.au)}
\address{\footnotemark[4]{} Department of Medical Physics and Biomedical Engineering, University College London, \\ Gower Street, London, WC1E 6BT, United Kingdom (b.treeby@ucl.ac.uk)}

\maketitle

\begin{history}
\received{(Day Month Year)}
\revised{(Day Month Year)}
\end{history}

\begin{abstract}
Model-based treatment planning and exposimetry for high-intensity focused ultrasound (HIFU) requires the numerical simulation of nonlinear ultrasound propagation through heterogeneous and absorbing media. This is a computationally demanding problem due to the large distances travelled by the ultrasound waves relative to the wavelength of the highest frequency harmonic. Here, the $k$-space pseudospectral method is used to solve a set of coupled partial differential equations equivalent to a generalised Westervelt equation. The model is implemented in C++ and parallelised using the message passing interface (MPI) for solving large-scale problems on distributed clusters. The domain is partitioned using a 1D slab decomposition, and global communication is performed using a sparse communication pattern. Operations in the spatial frequency domain are performed in transposed space to reduce the communication burden imposed by the 3D fast Fourier transform. The performance of the model is evaluated using grid sizes up to 4096 $\times$ 2048 $\times$ 2048 grid points distributed over a cluster using up to 1024 compute cores. Given the global nature of the gradient calculation, the model shows good strong scaling behaviour, with a speed-up of 1.7x whenever the number of cores is doubled. This means large-scale simulations can be distributed across high numbers of cores on a cluster to minimise execution times with a relatively small computational overhead. The efficacy of the model is demonstrated by simulating the ultrasound beam pattern for a HIFU sonication of the kidney.
\end{abstract}

\keywords{High Intensity Focused Ultrasound; Fourier Pseudospectral Methods; Westervelt Equation; Heterogeneous Media; Large-Scale Problems; Distributed Computing}


\section{\label{sec_introduction}Introduction}

High-intensity focused ultrasound (HIFU) is a non-invasive therapy in which a tightly focused beam of ultrasound is used to rapidly heat tissue in a localised region until the cells are destroyed \cite{Kennedy2003,TerHaar2007}. In recent years, HIFU has been used in clinical trials for the treatment of tumours in many organs, including the prostate, kidney, liver, breast, and brain \cite{Kennedy2003,Clement2004,Jolesz2008,Zhang2010a}. While the number of HIFU devices on the market continues to grow, one hurdle that currently prevents wider clinical use is the difficulty in accurately predicting the region of damaged tissue given a particular patient and set of treatment conditions. In principle, this could be calculated using appropriate acoustic and thermal models \cite{Paulides2013}. However, the modelling problem is both physically complex and computationally challenging. For example, the heterogeneous material properties of human tissue can cause the ultrasound beam to become strongly distorted \cite{Liu2005}, the exact values for the material properties and their temperature dependence are normally unknown \cite{Connor2002}, and the rate and mechanism for tissue damage are both temperature and cell specific \cite{Lepock2003}. 

The question of how best to model the physical interactions between ultrasound waves and biological tissue has been widely studied, and work in this area is ongoing. However, the problem that has attracted much less attention, but which is equally important, is the issue of computational scale. This arises because of two related factors. The first is that the generated acoustic pressures are of sufficient magnitude that the wave propagation is nonlinear. This causes the ultrasound waves to steepen as they propagate, which generates high-frequency harmonics of the source frequency (this is usually between 0.5 and 2 MHz for HIFU treatments where the transducer is positioned outside the body). At low focal intensities, nonlinear effects cause energy to be generated up to at least the 10$^\mathrm{th}$ harmonic \cite{Wojcik1995}. At very high focal intensities where strongly shocked waves are produced, as many as 600 harmonics might be required to model the focal heating accurately \cite{Khokhlova2010}. Thus the frequency content of the propagating ultrasound waves can be very broadband. 

The second factor is that the domain of interest encompassing the HIFU transducer and the treatment area is normally on the order of centimetres to tens of centimetres in each Cartesian direction. Compared to the size of the acoustic wavelength at the maximum frequency of interest, this equates to wave propagation over hundreds or thousands of wavelengths. To illustrate the scale of the problem, a list of representative domain sizes is given in Table \ref{TABLE_how_big}.  If the governing equations describing the HIFU field are solved on a uniform Cartesian grid where the grid spacing is defined to meet the Nyquist limit of two points per minimum wavelength, the resulting grid sizes can exceed 10$^{12}$ grid points.  If conventional finite difference schemes are used (which arguably is still the most common approach for time-domain modelling of broadband acoustic waves), the required grid sizes can be much greater. This is due to the large number of grid points per wavelength needed to avoid numerical dispersion over these length-scales. In many cases of practical interest, the grid sizes needed simply makes the simulations intractable.

\begin{table}[!t]
\tbl{Examples of possible domain sizes and frequency ranges encountered in high-intensity focused ultrasound (HIFU). The grid sizes are based on using a uniform Cartesian grid at the Nyquist limit of two points per minimum wavelength (PPMW) assuming a sound speed of 1500 m/s. The memory usage is based on storing a single matrix at the specified grid size in single precision (4 bytes per grid element).}
{\begin{tabular}{ccccc}
\toprule
Domain Size & Maximum Freq & Domain Size & Grid Size & Memory Per Matrix \\
(cm$^3$) & (MHz) & (wavelengths) & (at 2 PPMW) & (GB) \\
\toprule
5 $\times$ 5 $\times$ 5 & 5 & 333$^3$ & 667$^3$ &1.1 \\
& 10 & 667$^3$ & 1333$^3$ & 8.8\\
& 20 & 1333$^3$ & 2667$^3$ & 71\\
& 50 & 3333$^3$ & 6667$^3$ & 1100 \\
\midrule
10 $\times$ 10 $\times$ 10 & 5 & 667$^3$ & 1333$^3$ & 8.8\\
& 10 & 1333$^3$ & 2667$^3$ & 71\\
& 20 & 2667$^3$ & 5333$^3$ & 570\\
& 50 & 6667$^3$ & 13333$^3$ & 8800 \\
\midrule
20 $\times$ 20 $\times$ 20 & 5 & 1333$^3$ & 2667$^3$ & 71\\
& 10 & 2667$^3$ & 5333$^3$ & 570\\
& 20 & 5333$^3$ & 10667$^3$ & 4500 \\
& 50 & 13333$^3$ & 26667$^3$ & 71000 \\
\bottomrule
\end{tabular}}
\label{TABLE_how_big}
\end{table}

\begin{figure}[!t]
\centering
\includegraphics[width=0.76\textwidth]{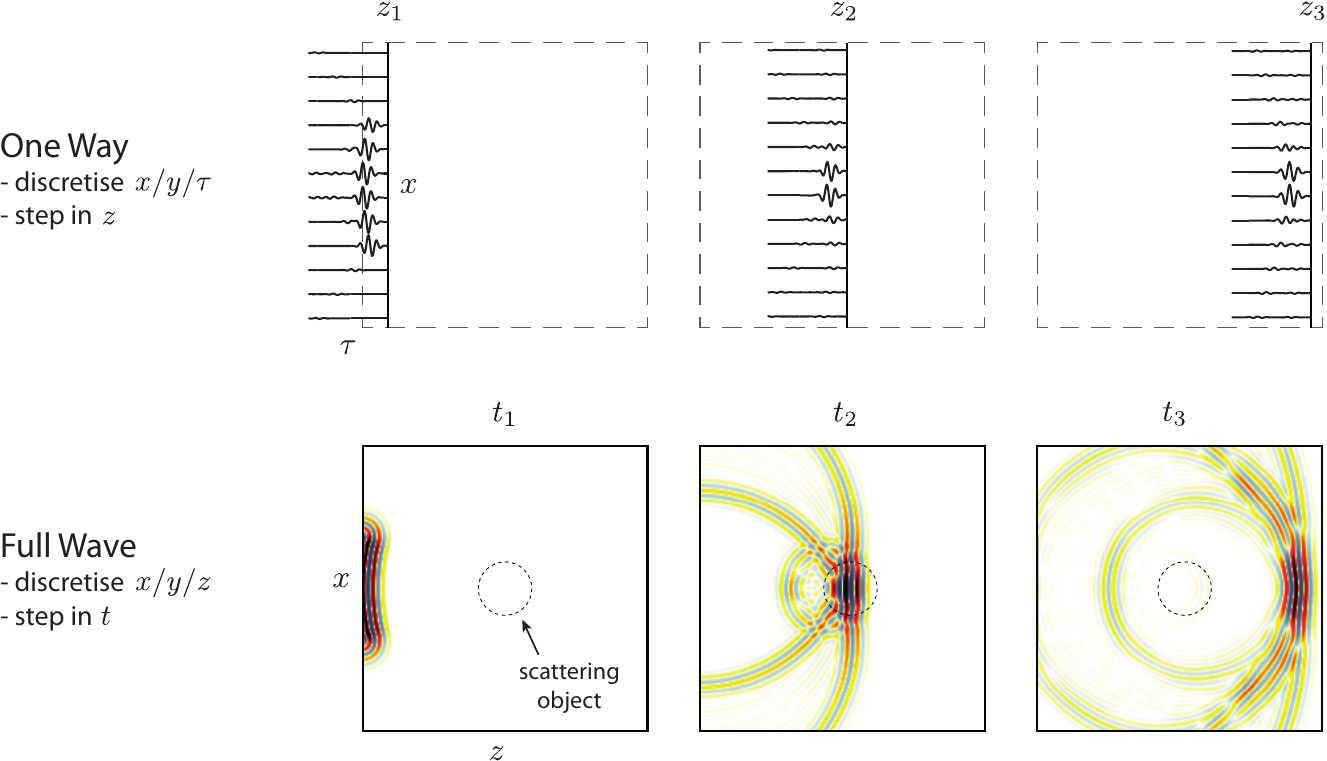}
\caption{Schematic illustrating the difference between one-way and full-wave ultrasound models. In one-way models, the governing equations are formulated in retarded time, and the domain is discretised in $x$, $y$, and $\tau$ (where $\tau$ is the retarded time variable). The simulation then progresses by propagating the time history of the wave-field from plane to plane in the $z$ dimension. In full-wave models, the domain is discretised in $x$, $y$, and $z$ and the simulation progresses by stepping through time. One-way models are typically more memory efficient, while full-wave models can account for heterogeneous material properties and reflected and scattered waves.}
\label{figure_one_way_vs_full_wave}
\end{figure}

To avoid the computational complexity of directly solving nonlinear acoustic equations in 3D, simplifying assumptions are normally made. In particular, one-way or evolution-type models have been very successful in simulating HIFU fields in homogeneous media \cite{Averkiou1999,Curra2000,Khokhlova2001,Yuldashev2011}. In one-way models, the governing equations are formulated in retarded time, and the domain is discretised in $x$, $y$, and $\tau$ (where $\tau$ is the retarded time variable) instead of $x$, $y$, and $z$. The simulation then progresses by propagating the time history of the wave-field from plane to plane in the $z$ dimension (this is illustrated in Fig.\ \ref{figure_one_way_vs_full_wave}). If a small time window is used, this approach can significantly reduce the amount of memory required for broadband simulations. For example, Yuldashev {\em et al.} used grid sizes in the $x$-$y$ plane with 10,000 $\times$ 10,000 grid points modelling up to 500 harmonics using a shared-memory computer with 32 GB of RAM \cite{Yuldashev2011}. The main restriction of one-way models is that a heterogeneous distribution of tissue properties cannot be included (except via the use of phase layers \cite{Yuldashev2010}), and scattered or reflected waves are not modelled. This means the significant distortion of HIFU beams that can occur in a clinical setting cannot be accounted for \cite{Liu2005}.

In addition to one-way models, several full-wave nonlinear ultrasound models in 3D have been reported based on the finite difference time domain (FDTD) method \cite{Connor2002,Pinton2009,Okita2011}. For example, Pinton {\em et al.} presented a solution to the Westervelt equation (a particular nonlinear wave equation) using a second-order-in-time, fourth-order-in-space finite difference scheme \cite{Pinton2009}. This was used to investigate the effects of nonlinear wave propagation on HIFU therapy in the brain \cite{Pinton2011a}. Simulations were run on 112 cores of a distributed cluster with grid sizes and run times on the order of $800 \times 800 \times 800$ and 32 hours. Okita {\em et al.} used a similar model to study HIFU focusing through the skull using 128 cores of a distributed cluster with grid sizes and run times on the order of $800 \times 600 \times 600$ and 2 hours \cite{Okita2011}. They also demonstrated excellent weak-scaling results for a benchmark using up to $3.45\times10^{11}$ grid points distributed across 98,304 cores on the K computer (a supercomputer
installed at the RIKEN Advanced Institute for Computational Science in Japan) \cite{Okita2014}. In another study, Pulkkinen {\em et al.} used a hybrid FDTD model to study transcranial focused ultrasound using grid sizes up to $1338 \times1363 \times 1120$ running on 96 cores of a distributed cluster with compute times on the order of 40 hours \cite{Pulkkinen2014a}. 

Computationally, FDTD schemes have excellent weak-scaling properties and good computational performance for a given grid size. However, as mentioned above, the significant drawback is that for large domain sizes, very dense grids are needed to counteract the accumulation of numerical dispersion. One way of reducing this computational burden is to compute spatial gradients using the Fourier pseudospectral method \cite{Hesthaven2007}. This eliminates the numerical dispersion that arises from the discretisation of spatial derivatives, and significantly reduces the number of grid points needed per acoustic wavelength for accurate simulations. Computational efficiency can be further enhanced by using the $k$-space pseudospectral method. This extends the pseudospectral method by exploiting an exact solution to the linearised wave equation to improve the accuracy of computing temporal gradients via a finite-difference scheme  \cite{Mast2001,Tabei2002}. This approach was first introduced in electromagnetics by Haber \cite{Haber1973}, and in acoustics by Bojarski \cite{Bojarski1982,Bojarski1985}. Since then, both pseudospectral and $k$-space schemes have been applied to linear \cite{Fornberg1987,Liu1999b,Mast2001,Tabei2002,Cox2007,Daoud2009,Tillett2009} and nonlinear \cite{Wojcik1998,Treeby2011a,Albin2012,Jing2012b,Treeby2012} ultrasound simulations in heterogeneous media for relatively modest grid sizes. Compared to FDTD schemes, the increase in accuracy arises due to the global nature of the basis functions (in this case complex exponentials) used to interpolate between the grid points. However, for large-scale problems where the numerical model must be implemented using a parallel computer with distributed resources, the global nature of the gradient calculation also introduces new challenges. This is due to the significant amount of global communication required between the compute cores \cite{Daoud2009}.

Another approach for minimising the grid size and number of time steps needed for accurate ultrasound simulations is the iterative nonlinear contrast source (INCS) method \cite{Verweij2009,Huijssen2010,Demi2011}. In this approach, terms describing the contributions of nonlinear effects and heterogeneous material parameters are re-formulated as contrast source terms. The resulting wave equation is then solved iteratively using Green's function methods \cite{Verweij2009}. While this works well for weakly non-linear fields, the requirement for storing the complete time history of the wave-field, and the evaluation of a 4D convolution at every time step makes it difficult to extend this approach to the large-scale problems encountered in HIFU.

Building on work by Tabei {\em et al.} and others \cite{Tabei2002}, we recently proposed an efficient full-wave nonlinear ultrasound model based on the $k$-space pseudospectral method \cite{Treeby2012}.  Here, we present an extension of this model for performing large-scale HIFU simulations on a distributed computer cluster with grid sizes up to 4096 $\times$ 2048 $\times$ 2048. A brief overview of the governing equations and the $k$-space pseudospectral model is given in Sec.\ 2. The software implementation and parallelisation strategies chosen for mapping the spectral gradient calculations onto a distributed computer cluster are then discussed in Sec.\ 3. In Sec.\ 4, performance and scaling results are presented for running simulations on up to 1024 cores. Results from a representative large-scale simulation of a HIFU treatment of the kidney are then presented in Sec.\ 5. Finally, summary and discussion are presented in Sec.\ 6.


\section{\label{sec_ultrasound_model}Nonlinear ultrasound model}

\subsection{Nonlinear governing equations}

For modelling the propagation of intense ultrasound waves in the human body, the governing equations must account for the combined effects of nonlinearity, acoustic absorption, and heterogeneities in the material properties (sound speed, density, acoustic absorption, and nonlinearity parameter). Following \cite{Treeby2012}, the required governing equations can be written as three-coupled first-order partial differential equations derived from the conservation laws and a Taylor series expansion for the pressure about the density and entropy 
\begin{align}
\frac{\partial \mathbf{u}}{\partial t} &= -\frac{1}{\rho_0} \nabla p + \mathbf{F} \enspace, &\text{(momentum conservation)} \nonumber \\
\frac{\partial \rho}{\partial t} &= -\rho_0 \nabla \cdot \mathbf{u} - \mathbf{u} \cdot \nabla \rho_0 - 2\rho \nabla \cdot \mathbf{u} + \mathrm{M} \enspace, &\text{(mass conservation)} \nonumber \\
p &= c_0^2 \left( \rho + \mathbf{d} \cdot \nabla \rho_0 + \frac{B}{2A}\frac{\rho^2 }{\rho_0} - \mathrm{L}\rho \right)\enspace. &\text{(pressure-density relation)} 
\label{EQ_governing_eqs_nonlinear_lossy}
\end{align}
Here $\vec{u}$ is the acoustic particle velocity, $\vec{d}$ is the acoustic particle displacement, $p$ is the acoustic pressure, $\rho$ is the acoustic density, $\rho_0$ is the ambient (or equilibrium) density, $c_0$ is the isentropic sound speed, and $B/A$ is the nonlinearity parameter which characterises the relative contribution of finite-amplitude effects to the sound speed. These equations account for cumulative nonlinear effects (nonlinear effects that build up over space and time) up to second order in the acoustic variables, equivalent to the Westervelt equation \cite{Westervelt1963,Hamilton2008}. All the material parameters are allowed to be heterogeneous. Two linear source terms are also included, where $\mathbf{F} $ is a force source term which represents the input of body forces per unit mass in units of N$\,$kg$^{-1}$, and $\mathrm{M}$ is a mass source term which represents the time rate of the input of mass per unit volume in units of kg$\,$m$^{-3}\,$s$^{-1}$. 

The nonlinear term in the mass conservation equation accounts for a convective nonlinearity in which the particle velocity affects the wave velocity. Using the linearised form of the equations given in Eq.\ \eqref{EQ_governing_eqs_nonlinear_lossy}, this term can be written in a number of different ways. Following \cite{Aanonsen1984}, the substitution of equations valid to first order in the acoustic variables into terms that are second order in the acoustic variables leads to third order errors, which can be neglected. This leads to
\begin{equation}
-2 \rho \nabla \cdot \mathbf{u} \approx \frac{2}{\rho_0}\rho\frac{\partial \rho}{\partial t} = \frac{1}{\rho_0}\frac{\partial \rho^2}{\partial t} \approx \frac{1}{\rho_0 c_0^4} \frac{\partial p^2}{\partial t} \enspace.
\end{equation}
In Eq.\ \eqref{EQ_governing_eqs_nonlinear_lossy}, the nonlinear term is written in the first form shown above as a spatial gradient of the particle velocity. This is significant because spatial gradients can be computed accurately using spectral methods, and don't require any additional storage. For comparison, the equivalent term in the Westervelt equation appears in the final form shown above as a temporal gradient of the square of the pressure \cite{Westervelt1963}. However, using this expression requires the use of a finite difference scheme and storage of the pressure field at previous time steps.

The operator $\mathrm{L}$ in the pressure-density relation in Eq.\ \eqref{EQ_governing_eqs_nonlinear_lossy} is an integro-differential operator that accounts for acoustic absorption that follows a frequency power law of the form $\alpha = \alpha_0 \omega^y$. This type of absorption has been experimentally observed in human soft tissues, where $y$ is typically between 1 and 2 \cite{Duck1990}. The operator has two terms both dependent on the fractional Laplacian and is given by \cite{Chen2004,Treeby2010a}
\begin{equation}
\mathrm{L} = \tau \frac{\partial}{\partial t} \left( -\nabla^2 \right)^{\tfrac{y}{2}-1} + \eta\left( -\nabla^2 \right)^{\tfrac{y + 1}{2} - 1} \enspace.
\label{EQ_L}
\end{equation}
Here $\tau$ and $\eta$ are absorption and dispersion proportionality coefficients given by $\tau = -2\alpha_0 c_0^{y-1}$ and  $\eta = 2 \alpha_0 c_0^y \tan \left( \pi y /2 \right)$, where $\alpha_0$ is the power law prefactor in $\mathrm{Np \,\left( rad/s\right)}^{-y}\, \mathrm{m^{-1}}$, and $y$ is the power law exponent. The two terms in $\mathrm{L}$ separately account for power law absorption and dispersion for $0 < y < 3$ and $y \ne 1$ \cite{Treeby2010a,Treeby2011d}.


\subsection{\label{sec_discrete_equations}Discrete equations}

Following Refs.\ \cite{Tabei2002} and \cite{Treeby2012}, the continuous governing equations given in the previous section can be discretised using the $k$-space pseudospectral method. If the mass conservation equation and the pressure-density relation given in Eq.\ \eqref{EQ_governing_eqs_nonlinear_lossy} are solved together, the $() \cdot \nabla \rho_0$ terms cancel each other, so they are not included in the discrete equations to improve computational efficiency. The mass and momentum conservation equations in Eq.\ (\ref{EQ_governing_eqs_nonlinear_lossy}) written in discrete form then become
\begin{subequations}
\label{EQ_discrete_linear}
\begin{align}
\frac{\partial}{\partial \xi} p^n &= \IFT{ i k_\xi \, \kappa \, e^{i k_\xi  \Delta \xi/2} \FT{ \vphantom{u_\xi^{n+1}} p^n} } \enspace, \label{EQ_discrete_linear_a} \\
u_\xi^{n+\tfrac{1}{2}} &= u_\xi^{n-\tfrac{1}{2}} - \frac{\DDelta t}{\rho_0} \frac{\partial }{\partial \xi} p^n + \DDelta t \,  \mathrm{F_{\xi}^n} \enspace, \label{EQ_discrete_linear_b} \\
\frac{\partial}{\partial \xi} u_\xi^{n+\tfrac{1}{2}} &= \IFT{ i k_\xi \, \kappa \, e^{-i k_\xi \Delta \xi/2} \FT{ u_\xi^{n+\tfrac{1}{2}} }}  \enspace, \label{EQ_discrete_linear_c} \\
\rho_\xi^{n+1} &= \frac{\rho_\xi^n - \DDelta t \rho_0 \frac{\partial }{\partial \xi}u_\xi^{n+\tfrac{1}{2}} }{  1 + 2\DDelta t  \frac{\partial }{\partial \xi}u_\xi^{n+\tfrac{1}{2}} } + \DDelta t \,  \mathrm{M_{\xi}^{n+\frac{1}{2}}}   \label{EQ_discrete_linear_d} \enspace.
\end{align}
\end{subequations}
Equations (\ref{EQ_discrete_linear_a}) and (\ref{EQ_discrete_linear_c}) are spatial gradient calculations based on the Fourier collocation spectral method, while (\ref{EQ_discrete_linear_b}) and (\ref{EQ_discrete_linear_d}) are update steps based on a $k$-space corrected finite difference scheme \cite{Tabei2002}. These equations are repeated for each Cartesian direction in $\mathbb{R}^n$ where $\xi = x$ in $\mathbb{R}^1$, $\xi = x, y$ in $\mathbb{R}^2$, and $\xi = x, y, z$ in $\mathbb{R}^3$.  Here $\FT{\ldots}$ and $\IFT{\ldots}$ denote the forward and inverse spatial Fourier transforms over $\mathbb{R}^n$, $i$ is the imaginary unit, $\DDelta t$ is the size of the time step, and $k_{\xi}$ represents the set of wavenumbers in the $\xi$ direction defined according to 
\begin{equation}
k_\xi = \left\{
\begin{array}{rl}
\left[-\tfrac{\mathrm{N}_\xi}{2}, -\tfrac{\mathrm{N}_\xi}{2} + 1, \ldots, \tfrac{\mathrm{N}_\xi}{2} -1 \right] \frac{2 \pi}{\DDelta \xi \, \mathrm{N}_\xi} & \qquad \text{if } \mathrm{N}_\xi \text{ is even} \\[1.5em]
\left[-\tfrac{(\mathrm{N}_\xi - 1)}{2}, -\tfrac{(\mathrm{N}_\xi - 1)}{2} + 1, \ldots, \tfrac{(\mathrm{N}_\xi - 1)}{2} \right] \frac{2 \pi}{\DDelta \xi \, \mathrm{N}_\xi} & \qquad \text{if } \mathrm{N}_\xi \text{ is odd} 
\end{array}
\right. \qquad.
\label{EQ_discrete_k_definition}
\end{equation}
Here $\mathrm{N}_\xi$ and $\DDelta \xi$ are the number and spacing of the grid points in the $\xi$ direction assuming a regular Cartesian grid. The $k$-space operator $\kappa$ in Eq.\ \eqref{EQ_discrete_linear} is used to correct for the numerical dispersion introduced by the finite-difference time step, and is given by $\kappa = \mathrm{sinc}\left( c_{\text{ref}} k \Delta t/2 \right)$, where $k = |\mathbf{k}|$ is the scalar wavenumber, and $c_{\text{ref}}$ is a single reference value of the sound speed (see discussion in Ref. \cite{Treeby2012} for further details).

\begin{figure}[!t]
\centering
\includegraphics{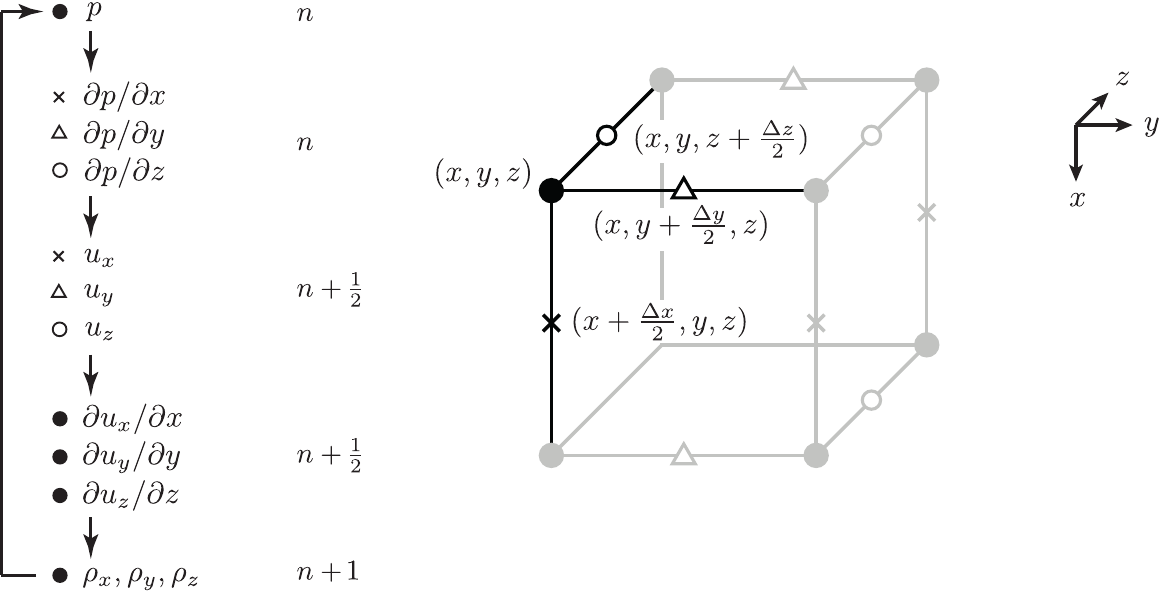}
\caption{Schematic showing the computational steps in the solution of the coupled first-order equations using a staggered spatial and temporal grid in 3D. Here $\partial p/ \partial x$ and $u_x$ are evaluated at grid points staggered in the $x$-direction (crosses), $\partial p/ \partial y$ and $u_y$ are evaluated at grid points staggered in the $y$-direction (triangles), and $\partial p/ \partial z$ and $u_z$ are evaluated at grid points staggered in the $z$-direction (open circles). The remaining variables are evaluated on the regular grid points (dots). The time staggering is denoted $n$, $n+\frac{1}{2}$, and $n+1$}
\label{figure_staggered_grid}
\end{figure}

The acoustic density and the mass source term (which are physically scalar quantities) are artificially divided into Cartesian components to allow an anisotropic perfectly matched layer (PML) to be applied. For the simulations presented here, Berenger's original split-field formulation of the PML is used \cite{Berenger1996} as described in \cite{Tabei2002}. The exponential terms $e^{\pm i k_\xi \Delta \xi /2}$ within Eqs.\ (\ref{EQ_discrete_linear_a}) and (\ref{EQ_discrete_linear_c}) are spatial shift operators that translate the result of the gradient calculations by half the grid point spacing in the $\xi$-direction. This allows the components of the particle velocity to be evaluated on a staggered grid. Note, the ambient density \smash{$\rho_0$} in Eq.\ (\ref{EQ_discrete_linear_b}) is understood to be the ambient density defined at the staggered grid points. The superscripts $n$ and ${n+1}$ denote the function values at current and next time points and $n-\tfrac{1}{2}$ and $n + \tfrac{1}{2}$ at the time staggered points. The time-staggering arises because the update steps, Eqs.\ (\ref{EQ_discrete_linear_b}) and (\ref{EQ_discrete_linear_d}), are interleaved with the gradient calculations, Eqs.\ (\ref{EQ_discrete_linear_a}) and (\ref{EQ_discrete_linear_c}). An illustration of the staggered grid scheme is shown in Fig.\ \ref{figure_staggered_grid}. 

The corresponding pressure-density relation written in discrete form is given by
\begin{equation}
p^{n+1} = c_0^2 \left( \rho^{n+1} + \frac{B}{2A} \frac{1}{\rho_0} \left( \rho^{n+1} \right)^2 - \mathrm{L_d} \right) \enspace,
\label{EQ_discrete_pressure_density}
\end{equation}
where the total acoustic density is given by $\rho^{n+1} = \sum_{\xi} \rho^{n+1}_\xi$, and $\mathrm{L_d}$ is the discrete form of the power law absorption term which is given by \cite{Treeby2010a}
\begin{equation}
\mathrm{L_d} = \tau \, \IFT{ k^{y-2} \, \FT{\frac{\partial \rho^n}{\partial t} }} + \eta \, \IFT{ k^{y-1} \, \FT{ \rho^{n+1} \vphantom{\sum{\frac{f}{f}}} }} \enspace.
\end{equation}
To avoid needing to explicitly calculate the time derivative of the acoustic density (which would require storing a copy of at least $\rho^n$ and $\rho^{n-1}$ in memory), the temporal derivative of the acoustic density is replaced using the linearized mass conservation equation $d\rho/dt = -\rho_0 \nabla \cdot \mathbf{u}$, which yields
\begin{equation}
\mathrm{L_d} = -\tau \, \IFT{ k^{y-2} \, \FT{ \rho_0 \sum \!\! \frac{}{}_\xi \frac{\partial}{\partial \xi} u_\xi^{n+\frac{1}{2}} }} + \eta \, \IFT{ k^{y-1} \, \FT{ \rho^{n+1} \vphantom{\sum{\frac{f}{f}}} }} \enspace.
\label{EQ_discrete_loss_operator}
\end{equation}
Further details about the formulation, stability, and accuracy of the $k$-space scheme can be found in Refs.\ \cite{Mast2001,Tabei2002,Cox2007,Treeby2010a,Treeby2012}.

\section{\label{sec_implementation}Implementation of the $k$-space pseudospectral method for distributed clusters}

\subsection{Overview}

The $k$-space and pseudospectral methods gain their advantage over finite difference methods due to the global nature of the spatial gradient calculations. This permits the use of a much coarser grid for the same level of accuracy. However, even using spectral methods, the computational and memory requirements for the nonlinear HIFU problems discussed in Sec.\ \ref{sec_introduction} are still considerable, and in most cases, significantly exceed the resources of a single workstation. In this context, the development of an efficient numerical implementation that partitions the computational cost and memory usage across a large-scale parallel computer is desired. However, the global nature of the gradient calculation, in this case using the 3D fast Fourier transform (FFT), introduces additional challenges for the development of an efficient parallel code. Specifically, while the FDTD method only requires small quantities of data to be exchanged between the processes responsible for adjacent portions of the domain, performing an FFT  requires a globally synchronising all-to-all data exchange. This global communication can become a significant bottleneck in the execution of spectral models. 

Fortunately, considerable effort has already been devoted to the development of distributed memory FFT libraries that show reasonable scalability up to tens of thousands of processing cores  \cite{Frigo2005,Pekurovsky2012}. These libraries have found particular utility in turbulence simulations where grid sizes up to 4096 $\times$ 4096 $\times$ 4096 have been used \cite{Yeung2012}. They have also been considered in previous implementations of linear $k$-space pseudospectral models using grid sizes up to 512 $\times$ 512 $\times$ 512 \cite{Daoud2009,Tillett2009}. The challenges for the current work were thus to determine how best to exploit these libraries within the context of the nonlinear $k$-space model, how to maximise the grid sizes possible for a given total memory allocation, how best to manage the generated output data, and how to maximise performance and scalability. 

The execution of the $k$-space pseudospectral model described in Sec.\ \ref{sec_ultrasound_model} can be divided into three phases: pre-processing, simulation, and post-processing. During the pre-processing phase, the input data for the simulation is generated. This involves defining the domain discretisation based on the physical domain size and  maximum frequency of interest, defining the spatially varying material properties (e.g., using a CT scan of the patient \cite{Schneider1996}), defining the properties of the ultrasound transducer and drive signal, and defining the desired output data (e.g., the peak positive pressure or time-averaged acoustic intensity in the region of the HIFU target \cite{TerHaar2011}). The simulation phase involves reading the input data, running the actual simulation following the discrete equations discussed in Sec.\ \ref{sec_discrete_equations}, and storing the output data. The post-processing phase involves analysing the (potentially large) output files and presenting this data in a human-readable form. Here, the discussion is focused primarily on the parallel implementation of the simulation phase. Some discussion of the pre- and post- processing stages is given in Sec.\ \ref{sec_application}. 

The discrete equations solved during the simulation phase are given in Eqs.\  \eqref{EQ_discrete_linear_a}-\eqref{EQ_discrete_linear_d} and \eqref{EQ_discrete_pressure_density}. Examining these equations, the complete set of data stored in memory during the simulation phase comprises:
\begin{itemize}
  \item Twenty-one real 3D matrices defined in the spatial domain, and three real and three complex 3D matrices defined in the spatial Fourier domain. These contain the medium properties at every grid point (sound speed, equilibrium density, absorption and non-linearity coefficients), the time-varying acoustic quantities (acoustic pressure, particle velocity, and acoustic density), the derivative and absorption operators, and temporary storage.
  \item Twenty real 1D vectors of various sizes
    defining the PML layer, the $k_\xi$ variables, the transducer drive signal, the indices of the grid points belonging to the transducer (referred to herein as the source mask), and the indices of the grid points where the output data will be collected (referred to herein as the sensor mask). 
  \item Approximately fifty scalar values defining the domain size, grid spacing, number of simulation time steps, and various simulation, control, and output flags. 
\end{itemize}
The operations performed on these datasets include 3D FFTs, element-wise matrix operations, injection of the source signal, and the collection of output data. 

The implementation of the discrete equations was written in C++ as an extension to the open-source k-Wave acoustics toolbox  \cite{Treeby2010,Treeby2012}. The standard message passing interface (MPI) was used to perform all interprocess communications, the MPI version of the FFTW library was used to perform the Fourier transforms \cite{Frigo2005}, and the  input/output (I/O) operations were performed using the hierarchical data format (HDF5) library. To maximise performance, the code was also written to exploit single instruction multiple data (SIMD) instructions such as streaming SIMD extensions (SSE). Further details of the implementation are given in the following sections.


\subsection{\label{sec_decomposition}Domain decomposition and the FFT}

To divide the computational domain across multiple interconnected nodes in a cluster, a one dimensional domain decomposition approach was used in which the 3D domain is partitioned along the $z$ dimension into 2D slabs. The slabs are then distributed over $P$ MPI processes, where each MPI process corresponds to one physical CPU core. The total number of processes is constrained by $P \leq \mathrm{N}_z$, where $\mathrm{N}_z$  is the number of grid points in the $z$ dimension (and thus the number of slabs). This decomposition approach was used as it is directly supported by the FFTW library, while other approaches, such as 2D partitioning, are not. 

\begin{figure}[!t]
\centering
\includegraphics[width=\textwidth]{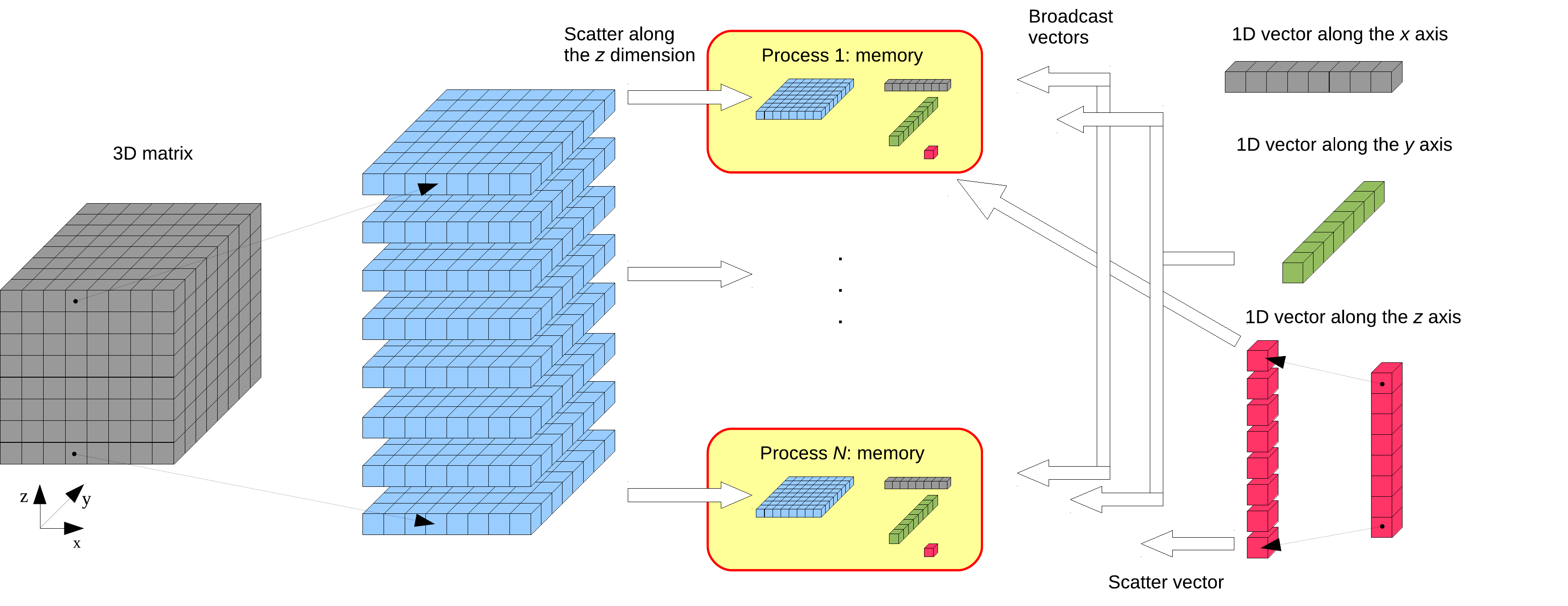}
\caption{Illustration of the 1D slab decomposition used to partition the 3D domain within a  distributed computing environment. The 3D matrices are partitioned along the $z$ dimension and distributed over $P$ MPI processes. 1D vectors oriented along the $x$ and $y$ dimensions are broadcast, while the vectors along the $z$ dimension are scattered. All scalar variables are broadcast and replicated on each process.} 
\label{figure_data_decomposition}
\end{figure}

Figure \ref{figure_data_decomposition} shows how the various spatial data structures are distributed to processes. For each 3D matrix there are a maximum of $\lceil \mathrm{N}_z/P \rceil$ 2D slabs stored on each process. For 1D quantities oriented along the $z$-axis, the data is partitioned and scattered over the processes in a similar manner. For 1D quantities oriented along either the $x$ or $y$ axis (and for scalar quantities), the data is broadcast and replicated on every process. The exception is for the source and sensor masks, which list the grid indices where the input data is defined and where the output data is collected. These are distributed such that individual processes are assigned the portion of the list that corresponds to parts of the source and/or sensor that fall within its local section of the domain. As the source and sensor masks do not usually cover the whole domain, many processes are likely to receive no source or sensor related data.

\begin{figure}[!t]
\centering
\includegraphics[scale=0.85]{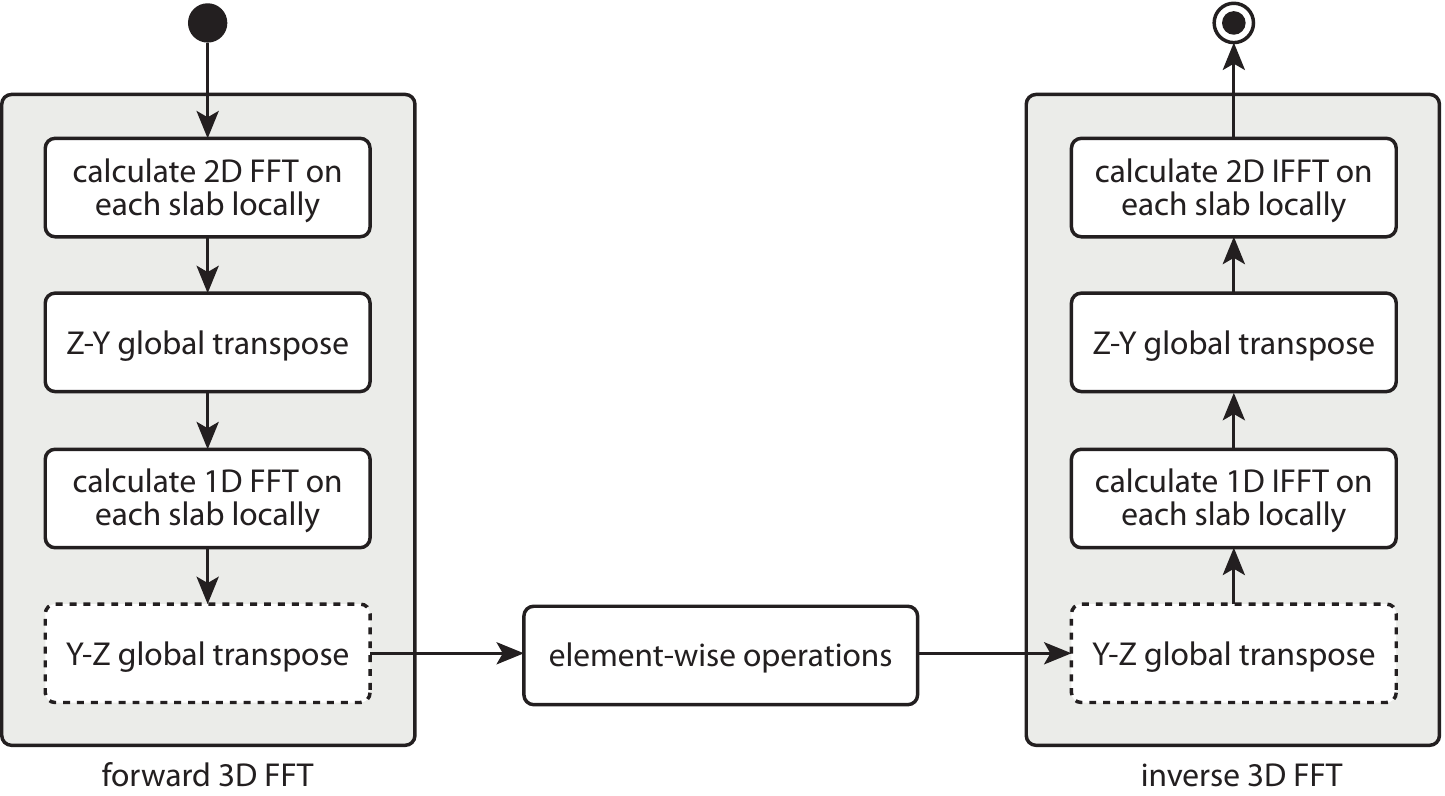}
\caption{Chain of operations needed to calculate spatial gradients using the Fourier pseudospectral method. First, the 3D forward FFT is calculated, then element-wise operations within the spatial frequency domain are applied. Finally the result is transformed back to the spatial domain using an inverse 3D FFT. The transposes depicted in the dashed boxes can be omitted if the element-wise matrix operations are performed in the transposed domain.} 
\label{figure_fft_chain}
\end{figure}

The calculation of the spatial gradients following Eqs.\ \eqref{EQ_discrete_linear_a} and \eqref{EQ_discrete_linear_c} involves  performing a 3D FFT, followed by one or more element-wise matrix operations, followed by an inverse 3D FFT. Using the domain decomposition scheme outlined above, each 3D FFT is executed by first performing a series of 2D FFTs in the $xy$ plane (i.e., on each slab) using local data. This is then followed by an all-to-all global communication to perform a $z \leftrightarrow y$ transposition. This step is necessary as the FFT can only be performed on local data (it cannot stride across data belonging to multiple processes).  The global transposition is then followed by a series of 1D FFTs performed in the transposed $z$ dimension, followed by another global transposition from $y \leftrightarrow z$ to return the data to its original layout. This chain of operations is illustrated in Figure \ref{figure_fft_chain}. In this decomposition, the main performance bottleneck is the two global transpositions required per FFT. 

Examining Fig.\ \ref{figure_fft_chain}, it is apparent that  the last global $y \leftrightarrow z$ transposition of the forward FFT is paired with an identical but reverse transposition immediately after the element-wise operations. As the intervening operations are independent of the order of the individual elements, it is possible to eliminate these two transpositions such that operations in the spatial frequency domain are performed in transformed space \cite{Frigo2012a}. This has a significant effect on performance, with compute times reduced by 35-40\% depending on the number of processes used. Note, in this case, variables defined in the spatial frequency domain must instead be partitioned in transformed space along the $y$-dimension, with the total number of MPI processes  constrained by $P \leq \text{min} \left( \mathrm{N}_y, \mathrm{N}_z \right)$. To avoid having idle processes during calculations involving either regular or transposed data, the number of processes $P$ should ideally be a chosen to be a divisor of both $\mathrm{N}_y$ and $\mathrm{N}_z$.

As the input data to the forward FFT is always real, the output of the Fourier transform in the first dimension is symmetric. Computational efficiency can thus also be improved by using the real-to-complex and complex-to-real FFTW routines which only calculate and return the unique part of the transform. To further improve performance, the element-wise matrix operations executed in between these transforms, as well as those defined in Eqs.\ \eqref{EQ_discrete_linear_b} and \eqref{EQ_discrete_linear_d}, were merged into compute kernels optimised to maximise temporal data locality and performance. In some cases, the latter involved manually inserting calls to SIMD intrinsic functions into loop structures that the compiler was unable to vectorize otherwise.

\subsection{Input and output files}

The open-source HDF5 library was chosen to manipulate the input and output files because of its ability to organise complex datasets in heterogeneous computing environments. This library is available on most supercomputers and is also supported by numerical computing languages such as MATLAB and Python which is useful for pre- and post- processing. The HDF5 library provides two interfaces: serial and parallel. The serial interface targets shared memory computers and assigns a single thread to be responsible for all I/O. This interface was used to generate the input files during the pre-processing phase using a single large-memory node. The parallel interface targets clusters and essentially provides a user-friendly wrapper to the low level MPI-I/O library. This allows multiple processes to share the same file and read or write to it simultaneously. This interface was used during the simulation phase as it provides much higher I/O performance than either serialised accesses or master-slave architectures (where a single process serves all I/O requests). The parallel HDF5 interface also allows for two different I/O access patterns: independent and collective. In most cases, the collective mode is preferred as it enables the HDF5 runtime to rearrange and reshape data from different parallel processes before writing it to disk. This mode was used for all I/O operations during the simulation phase, the only exception being when writing scalar values to the output file.

Within the input and output files, all datasets were stored as three dimensional arrays in row-major order, with dimensions defined by the tuple $(\mathrm{N}_x, \mathrm{N}_y, \mathrm{N}_z)$. For scalars and 1D and 2D arrays, the unused tuple dimensions were set to 1. For example, scalar variables were stored as arrays of dimension $(1, 1, 1)$, 1D vectors oriented along the $y$-axis were stored as arrays of dimension $(1, \mathrm{N}_y, 1)$, etc. For datasets containing complex data, the lowest used dimension was doubled and the data stored in interleaved format. Additional information about the simulation (for example, the control flags and parameters) were stored within the file header.

To maximise I/O performance, the datasets within the input and output files were stored in a chunked layout, where each dataset is divided into multiple blocks or chunks that can be accessed individually. This is particularly useful for improving the throughput of partial I/O, where only portions of a dataset are accessed at a time. The use of chunking also provides a convenient way to overcome one of the current MPI limits, namely the 2 GB maximum size of a message (this is due to the routine headers in the MPI standard being defined using signed integers). Without chunking, this limit is easy to exceed, particularly during MPI gather operations where hundreds of small messages are aggregated. In addition to partial I/O, chunking enables the use of on-the-fly data compression. This is particularly beneficial for the input file, as there are often large areas of the domain with similar material properties, for example, the layer of water or coupling medium between the transducer and patient. These homogeneous regions give rise to good compression ratios, and thus reduce the amount of communication necessary when loading the input file. However, while both the serial and parallel HDF5 interfaces can read compressed datasets, only the serial interface can write such datasets. Thus compression was only used for the input files.

For the 3D datasets, the chunk size was chosen to be a single 2D slab matching the decomposition discussed in Sec.\ \ref{sec_decomposition}. This is the smallest portion of data read by each MPI process.  Each slab is only ever accessed by one process at a time, and is usually a reasonable size in terms of I/O efficiency. For the 2D and 1D quantities, the data was divided into chunks of fixed size along the lowest used dimension. A chunk size of 8MB was found to give reasonable I/O performance. Note, it is possible to further tune the chunk size for different size datasets to maximise I/O performance on a given platform. This can be useful on parallel cluster file systems that use hundreds of disk drives spread over many nodes in conjunction with complex internode communication patterns.


\subsection{Simulation stages}

\begin{figure}[!t]
\centering
\includegraphics[scale=0.85]{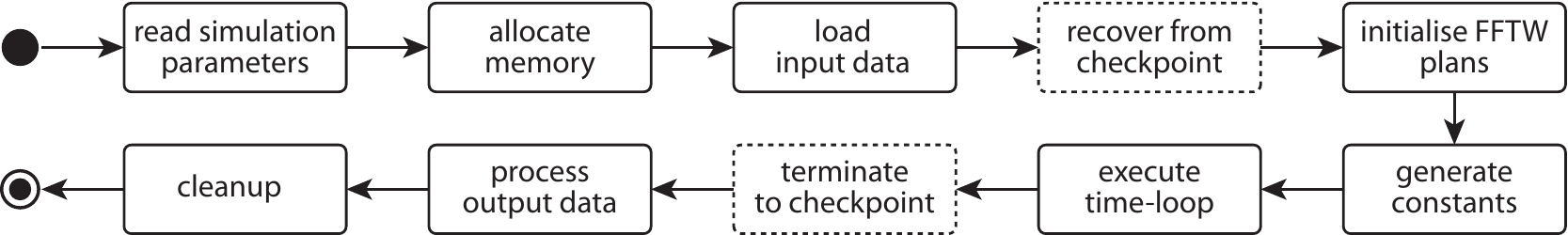}
\caption{Overview of the stages within the simulation phase of the parallel implementation of the $k$-space pseudospectral model.}
\label{figure_program_flow}
\end{figure}

An overview of the stages within the simulation phase of the parallel implementation of the $k$-space pseudospectral model is shown in Fig.\ \ref{figure_program_flow}. When the simulation begins, each MPI process parses the command line parameters specifying the location of the input and output files. The input file is then opened and parameters such as the domain size and the number of time steps are loaded. This is performed using a HDF5 broadcast read. In this operation, all processes select the same item to be read from the file. The HDF5 library then joins all I/O requests such that the data is only physically read from disk once and then broadcast over all processes using an MPI routine. Next, the 1D decomposition of the domain is calculated as discussed in Sec.\ \ref{sec_decomposition}, and memory for the various simulation and temporary quantities is allocated. Note, the use of SIMD instructions imposes several requirements on the data layout and its memory alignment. First, all multidimensional matrices must be stored as linear arrays in row-major order. Second, complex quantities must be stored in interleaved format. Third, data structures must be allocated using an FFTW memory allocation routine that ensures they are aligned on 16B boundaries.

After memory allocation, the contents of the input file are loaded and distributed over all processes. The 3D datasets (e.g., the medium properties) are loaded in chunks, with each process identifying its local portion of the domain and invoking a collective HDF5 read operation. The 1D vectors that are scattered are loaded in a similar fashion, while 1D vectors and scalar values that are replicated are read using a HDF5 broadcast read. The source and sensor masks are loaded in chunks and broadcast to all processes, with each process extracting the grid indices that fall within its portion of the domain. Once the distribution of the source mask is known, the drive signal for the transducer is then sent to the relevant processes. Finally, the distribution of the sensor mask is used to allocate appropriate buffers to store the output data. These buffers are eventually flushed into the output file.

Next, execution plans for the FFT are generated. This step is required by FFTW prior to performing the first transform of a particular rank, size, and direction. It involves the library trying several different FFT implementations with the objective of finding the fastest execution pathway on the current hardware \cite{Frigo2005}. Consideration is given to many architectural aspects including the CPU, the memory system, and the interconnection network. Depending on the domain size, this step can take a considerable amount of time. However, as it is only executed once and there are many thousands of FFTs in a typical simulation, the benefit of having a fast implementation is usually significant (this is discussed further in Sec.\ \ref{sec_strong_scaling}). Unfortunately, while it is possible to save plans between code executions, there is usually little benefit in doing so. This is because the plans are problem size specific, and depend on the number and distribution of cores within the parallel system being used. On shared clusters in particular, the distribution of cores is likely to change between runs according to the cluster utilisation and management system.

After the FFT plans are generated, simulation constants such as the $k$-space operator ($\kappa$) and the absorption parameters ($\tau$, $\eta$, $k^{y-2}$, $k^{y-1}$) are generated. The simulation time-loop then begins following the equations defined in Eqs.\ \eqref{EQ_discrete_linear_a}-\eqref{EQ_discrete_linear_d} and \eqref{EQ_discrete_pressure_density}. For each time step, there are six forward and eight inverse FFTs. There are two fewer forward FFTs as the three spatial components in Eq.\ \eqref{EQ_discrete_linear_a} share the same $\FT{ p^n }$. The source injection is implemented as an operation that updates the value of the acoustic pressure or velocity at the relevant grid points within the domain. In a similar vein, the output data is collected by storing the acoustic parameters at the grid points specified by the sensor mask. Aggregate quantities (e.g., the peak positive pressure) are kept in memory until the simulation is complete, while time-varying quantities are flushed to disk at the end of each time step using a collective HDF5 write routine. When the time-loop is complete, performance data related to the  execution time and memory usage is also written to the output file. The output file is then closed and the memory cleared.

When running larger simulations, a checkpoint-restart capability is also used. This allows large simulations to be split into several phases, which is useful for staying within wall-clock limitations imposed on many clusters, in addition to providing a degree of fault tolerance. During checkpointing, an additional HDF5 output file is created which stores the current values of the time-varying acoustic quantities as well as the aggregated output quantities. Restart is performed in the same way as a new simulation, however, after the input file has been loaded, the checkpoint file is opened and the acoustic quantities and the aggregated values are restored. 


\section{Performance evaluation}

\subsection{\label{sec_benchmark}Benchmark platform}

To evaluate the utility of the implemented $k$-space pseudospectral model in the context of large-scale nonlinear ultrasound simulations for HIFU, a number of performance metrics were investigated. These included the strong and weak scaling properties of the code, the absolute simulation time and cost, and the effect of the underlying hardware architecture. The tests were performed using the VAYU supercomputer run by the National Computational Infrastructure (NCI) in Australia. This system comprises 1492 nodes, each with two quad-core 2.93GHz Intel Nehalem architecture Xeon processors and 24GB of memory configured in a non-uniform memory access (NUMA) design.  The compute nodes are connected to the cluster via an on-board infiniband interface with a theoretical bandwidth of 40Gb/s, while the interconnection network consists of four highly integrated 432-port infiniband switches. This relatively simple network design reduces the impact of process placement on interprocess network bandwidth and latency. The I/O disk subsystem is physically separated from the compute nodes. The Lustre parallel distributed file system is used to manage 832 TB of disk space distributed over 1040 disk drives. This configuration offers very high bandwidth, however, latency for small I/O transactions (e.g., storing less than 1MB of data) can be relatively high. Also, as the I/O subsystem is shared amongst all users, notable performance fluctuations can occasionally be observed.

VAYU runs a highly modular CentOS Linux distribution which offers a number of development environments (compilers and libraries) that can be loaded on the fly based on application requirements.  The cluster is managed by the OpenPBS batch queuing system. A job for execution is wrapped with its compute requirements such as the number of compute cores, amount of main memory, disk space, execution time, and software modules. The batch queuing system then schedules the job for execution based on the priority and actual system workload. The resources granted to the jobs are always dedicated. After the job completes, the user is charged a number of service units, where one service unit corresponds to one core-hour used (when running on the standard priority queue). For example, if a job takes 2 hours of wall-clock time running on 1024 cores, the user would be charged 2048 service units.

The performance metrics were evaluated using a common set of benchmarks. The benchmark set was designed to cover a wide range of domain sizes, from small simulations that can be run on desktop systems, up to large-scale simulations that approach the limits of what is currently feasible using a supercomputer. The simulations accounted for nonlinear wave propagation in an absorbing medium in which all material properties were heterogeneous. The grid sizes increased from $2^{24}$ grid points ($256 \times 256 \times 256$) to $2^{34}$ grid points ($4096 \times 2048 \times 2048$), where each successive benchmark was doubled in size, first by doubling the grid size in the $x$ dimension, then the $y$ dimension, then the $z$ dimension, and so on. The benchmarks were executed on VAYU using different numbers of compute cores ranging from 8 (one node) to 1024 (128 nodes) in multiples of 2. The minimum and maximum number of cores used for a particular grid size were limited by the available memory per core from the bottom (24 GB per node or 3 GB per core), and the size of the simulation domain from the top ($P \leq \text{min} \left( \mathrm{N}_y, \mathrm{N}_z \right)$). Compute times were obtained by averaging over 100 time-steps.

\begin{figure}[!th]
\centering
\includegraphics{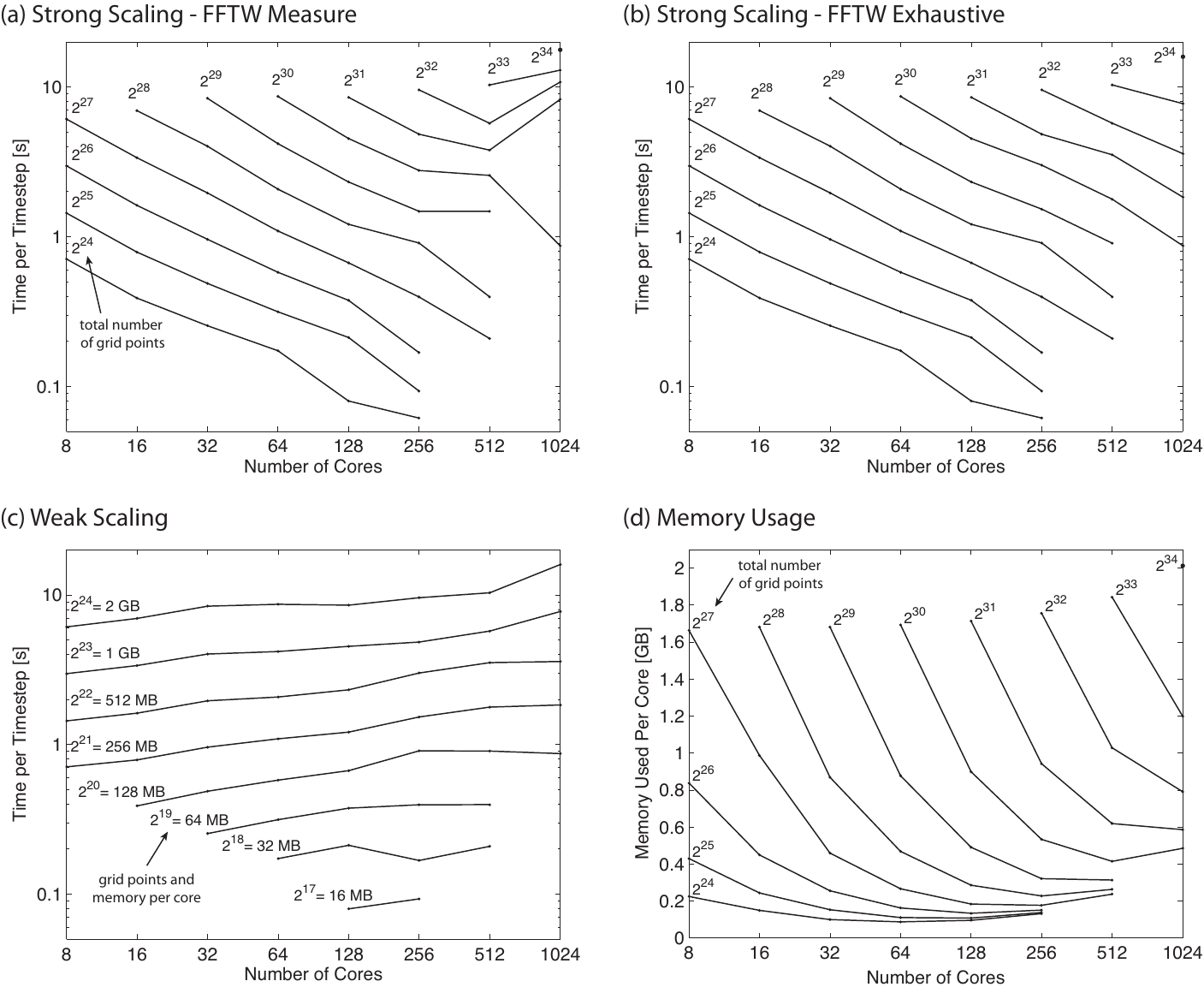}
\caption{Performance results for nonlinear ultrasound simulations in heterogeneous and absorbing media. (a), (b) Strong scaling showing the variation of the execution time with the number of CPU cores for a fixed grid size. The domain size varies from $2^{24}$ ($256^{3}$) grid points to $2^{34}$ ($4096 \times 2048 \times 2048$) grid points. The plots clearly show the difference in strong scaling between \texttt{FFTW\_MEASURE} and \texttt{FFTW\_EXHAUSTIVE} flags used for planning the execution of FFTs and related global communications. (c) Weak scaling showing the potential of more cores to solve bigger problems in the same wall-clock time. The curves show the working dataset size per core growing from 16 MB ($2^{17}$ grid points) up to 2 GB ($2^{24}$ gird points). (d) Memory usage per core for the strong scaling results given in part (b). }
\label{figure_scaling}
\end{figure}


\subsection{\label{sec_strong_scaling} Strong scaling}

Strong scaling describes how the execution time changes when using an increasing number of compute cores for a fixed problem size. Ideally, whenever the number of compute cores is doubled, the execution time should reduce by a factor of 2. However, in practice, all but embarrassingly parallel applications include some level of overhead that causes them to eventually reach a point where increasing the number of compute cores does not reduce the execution time, or can even make it worse. 

Figures \ref{figure_scaling}(a) and \ref{figure_scaling}(b) show the strong scaling results obtained for the eleven different benchmark problem sizes using two different FFTW planing flags, \texttt{FFTW\_MEASURE} and \texttt{FFTW\_EXHAUSTIVE}. These flags determine the extent to which FFTW attempts to find the best FFT implementation during the plan generation phase. The flag \texttt{FFTW\_MEASURE} is the default, while the flag \texttt{FFTW\_EXHAUSTIVE} triggers a much more extensive search covering the full range of options implemented by the FFTW library. When using  the \texttt{FFTW\_MEASURE} flag, the results show almost ideal strong scaling for domain sizes up to $2^{31}$ grid points ($2048 \times 1024 \times 1024 $) and core counts up to 256. Within this region, all the curves have approximately equal gradients corresponding to a speed-up of roughly 1.7x when doubling the number of cores. Beyond 256 cores, acceptable scaling is obtained for many cases, however, there is virtually no improvement in execution time for grid sizes of $2^{29}$ and $2^{30}$ grid points. Further increasing the core count to 1024 leads to significantly worse performance, with the exception of the grid size of $2^{30}$ grid points ($1024 \times 1024 \times 1024 $) where, interestingly, the performance increases markedly. 

The erratic performance of FFTW at large core counts when using the \texttt{FFTW\_MEASURE} flag is due to differences in the communication pattern chosen during the planning phase. Specifically, using the integrated performance monitoring (IPM) profiler, it was found that in most cases, the global transposition selected by FFTW was based on a simple all-to-all MPI communication routine which exchanges $P(P-1)$ messages amongst $P$ compute cores. This becomes increasingly inefficient when the message size drops into the tens of kB range and the number of messages rises into the millions. This is the case for the very large simulations on high core counts shown in Fig.\ \ref{figure_scaling}(a). However, FFTW also includes other communication algorithms, including one that uses a sparse communication pattern. This routine was selected by \texttt{FFTW\_MEASURE} for the simulation with $2^{30}$ grid points and 1024 cores, leading to significant performance gains. 

The sparse communication pattern used by FFTW combines messages local to the node into a single message before dispatch. In the case of VAYU, this means that 8 messages are combined together within each node before sending to other nodes. This decreases the number of messages by a factor of 64 and also increases the message size. When using  \texttt{FFTW\_EXHAUSTIVE}, the sparse communication pattern was always selected. As shown in Figure \ref{figure_scaling}(b), the impact of this on performance is considerable. Almost all anomalies are eliminated, and the scaling remains close to ideal over the whole range of grid sizes and cores counts considered. This is consistent with profiling data, which revealed that 30-40\% of the execution time was associated with communication,  another 30-40\% due to FFT operations, and the remaining 15-20\% due to element wise matrix operations. These results indicate that given large enough grid sizes, reasonable scalability to even larger numbers of compute cores should be possible. Small deviations in the scaling results are likely to be caused by the degree of locality in the process distribution over the cluster, as well as by process interactions when performing I/O operations.

In addition to the benchmark set considering nonlinear wave propagation in a heterogeneous and absorbing medium, additional benchmarks were performed considering (1) nonlinear wave propagation in a homogeneous and absorbing medium, and (2) linear wave propagation in a homogeneous and lossless medium.  The strong-scaling results for both cases were similar to those given above. This is not surprising given that the FFT is the dominant operation in all test cases. In terms of absolute performance, the results for case (1) benefit from both a reduced memory footprint and a reduced memory bandwidth requirement. Specifically, because the medium is homogeneous, the medium parameters can be described using scalar variables that are easily cached, rather than large 3D matrices that are usually read from main memory each time they are used. This reduces the memory requirements by approximately 30\% and the execution times by 5-10\%. For case (2), an additional advantage comes from a reduction in the number of FFTs performed within each time step (ten instead of fourteen). This reduction is mirrored almost exactly in the observed execution times, which were typically 30\% less than the execution times for an absorbing medium.


\subsection{Weak scaling}
Weak scaling describes how the execution time changes with the number of compute cores when the problem size per core remains fixed. In the ideal case, doubling the number of compute cores should enable a problem twice as big to be solved in the same wall-clock time. Since the computational work per core is constant, overhead increases only if the communication pattern becomes more complicated as the problem size is increased. Figure \ref{figure_scaling}(c) shows the weak scaling results for the benchmark set defined in Sec.\ \ref{sec_benchmark} when using the \texttt{FFTW\_EXHAUSTIVE} planing flag. Despite the fact that the cost of the underlying all-to-all communication step grows as the square of the number of processes, the graph shows relatively good weak scaling performance. The trends suggest that simulations with even larger grid sizes could be solved with reasonable efficiency using higher numbers of compute cores.


\subsection{Memory usage}

Figure \ref{figure_scaling}(d) shows how the memory usage per core changes in-line with the strong scaling results given in Figure \ref{figure_scaling}(b). Whenever the number of cores is doubled, the domain is partitioned into twice as many 1D slabs. Theoretically, the memory required by each core should be halved. However, in practice, there is an overhead associated with the domain decomposition that grows with the number of cores, and eventually becomes dominant. This behaviour is clearly observed for almost all the simulation sizes in the benchmark set. The overhead is comprised of several components, each with a different origin.

First, the MPI runtime allocates a significant amount of memory for communication buffers. These buffers serve as a reservation station for incoming messages (like a platform for a train), such that when a message arrives, there is always a free place in memory where the data can be immediately housed using a direct memory transfer from the network interface. When the MPI library is not restricted, it allocates as many communication buffers for each process as there are distinct messages to be received (one for each sender). Moreover, if the sparse communication pattern is used, additional scratch space must be allocated where smaller messages can be combined and separated on send and receive. For simulations using high process counts where the memory used per core for the local partition is quite low (tens of MB), the memory allocated for MPI communication buffers can become the  dominant component.

In addition to communication buffers, when the domain is partitioned into more parts over an increasing number of processes, the total memory consumed by locally replicating scalar variables and 1D arrays is increased. Moreover, additional memory is needed for storing the code itself. While the size of the compiled binary file is only on the order of 20 MB, a private copy is needed for every process. Thus, when 1024 processes are used, storing the code consumes more than 20 GB of memory. For small simulations, this can become a significant portion of the total memory consumption.


\subsection{Simulation cost and execution time}

For the user, another important metric that must be considered when running large-scale simulations is their financial cost. For this purpose, the simulation cost is defined as the product of  the wall-clock execution time and the number of compute cores used. On VAYU, this cost is expressed in terms of service units (SUs). These are directly related to the number of core-hours used scaled by a few other factors such as the priority the user assigns to the job. SUs represent an accountant's view on the parallel economy. On a large shared parallel computer, each user is typically allocated a share of the resource. On VAYU this corresponds to a certain number of SUs per quarter, and it is left to the user to determine how best to use this allocation. As scaling is never ideal, the more compute cores assigned to a job, the higher the effective cost. However, for time critical problems, using the highest possible number of cores may still be desirable.

\begin{figure}[!t]
\centering
\includegraphics[width=\textwidth]{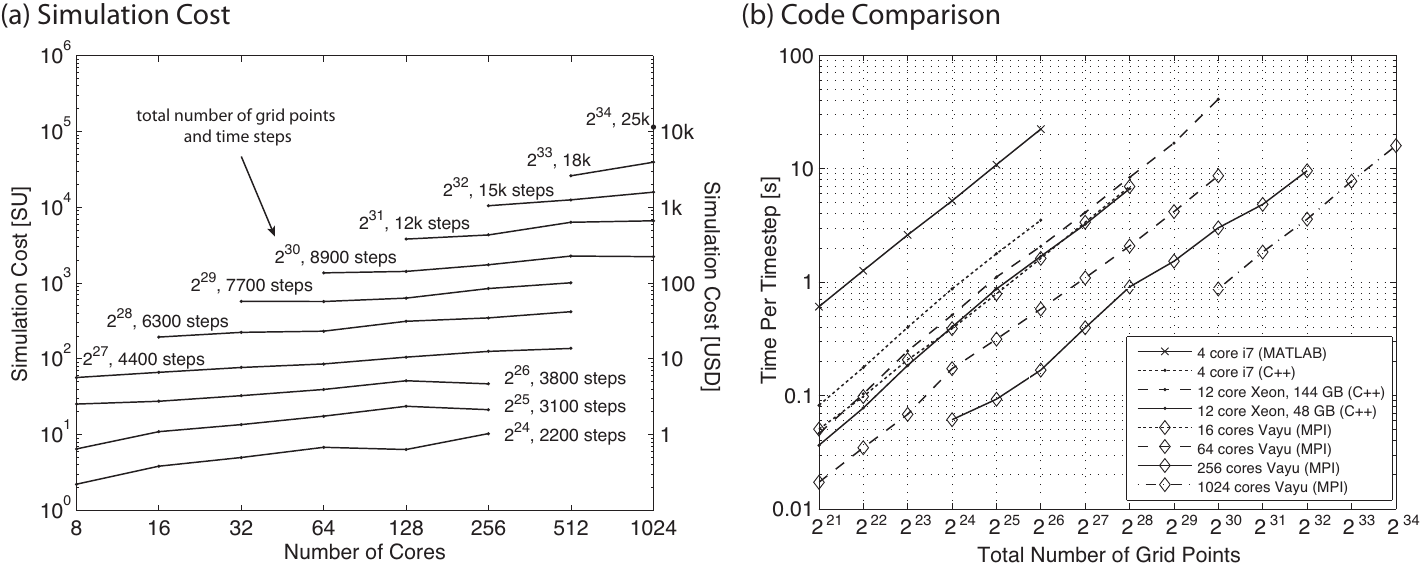}
\caption{(a) The simulation cost in terms of service units (core-hours) displayed on the left y axis and USD on the right y axis. The number of time steps is derived from the time the wave needs to propagate along the diagonal length of the domain. (b) Execution time per timestep running different implementations of the $k$-space pseudospectral model on different machines.}
\label{figure_cost_comparison}
\end{figure}

Figure \ref{figure_cost_comparison}(a) shows the anticipated total simulation cost for the grid sizes in the benchmark set against the number of compute cores used. As the grid size increases, more time steps are also necessary to allow the ultrasound waves to propagate from one side of the domain to the other. Using the diagonal length of the domain and assuming a Courant-Friedrichs-Lewy (CFL) number of 0.2, the number of time steps grows from 2200 for a grid size of $256 \times 256 \times 256$ to 25,000 for a grid size of $4096\times 2048 \times 2048$ grid points. The results in Fig.\ \ref{figure_cost_comparison}(a) show that for a given grid size, the simulation cost remains fairly constant as a function of core count, with the ratio between the highest and lowest costs always less than 2 (this is to be expected given the strong-scaling results). 

NCI, the owners of VAYU, charge US\$0.1 per SU to commercial projects. Using this value, the approximate cost of each simulation in US dollars is also shown in Fig.\ \ref{figure_cost_comparison}(a). If run to completion, the largest simulation performed would take around 4.5 days on 1024 compute cores and would cost slightly over US\$10,000. At this point in time such a large simulation is clearly not routine. However, VAYU is now a relatively old system, and with continued price performance trends it is not impossible to see the cost of such a simulation dropping to a few hundred dollars within the next few years.


\subsection{Comparison of different architectures }

Thus far, only the performance of the MPI implementation of the $k$-space pseudospectral model running on the VAYU cluster has been considered. For comparison, Fig.\ \ref{figure_cost_comparison}(b) illustrates the execution times per time step for two equivalent models implemented in (a) MATLAB and (b) C++ but parallelised for shared-memory architectures using OpenMP rather than MPI (some of the details of this implementation are discussed in \cite{Jaros2012}). For these comparisons, the benchmark set was extended to include smaller grid sizes starting at $2^{21}$ grid points (128 $\times$ 128 $\times$ 128). Three different computer configurations were considered: (1) a desktop machine with 12GB of RAM and a single Intel Core i7 950 quad-core CPU running at 3.2GHz, (2) a server with 48 GB of RAM and two Intel Xeon X5650 hex-core CPUs running at 2.66GHz, and (3) an identical server with 144 GB of RAM. The two server configurations differ in that the memory controller is not able to operate at 1333MHz when serving 144GB of RAM and drops its frequency to 800MHz. The effect of this is to reduce the memory bandwidth by about 33\%. 

Each line in Fig.\ \ref{figure_cost_comparison}(b) represents a different combination of computer configuration and implementation. The slowest performing combination is the MATLAB implementation running on the quad-core i7 system (this was the starting point for the development of the C++ codes \cite{Treeby2010}). Using the shared-memory C++ implementation on this machine reduces the execution time by a factor of about 8. Moving to the 12-core Xeon systems more than doubles this factor, although there are clear differences between the 48GB and 144GB configurations due to the fact that the model is memory bound. In comparison, to match the performance of the shared memory code running on the 12-core Xeon system, the MPI version of the code requires 16-cores on VAYU. The higher core-count needed by the MPI version of the code for equivalent performance reflects the fact that a non-trivial overhead is incurred since all communications between cores must pass through the MPI library (and in many cases, over the network). Of course, the benefit of the MPI implementation is the possibility of running the code over multiple nodes on a cluster, enabling both much larger domain sizes and much faster execution times. For example, for a grid size of $2^{26}$ grid points, the MPI code running on 256 cores is approximately two orders of magnitude faster than the MATLAB implementation on the quad-core i7 system, and one order of magnitude faster than the shared memory C++ implementation running on the 12-core Xeon systems. 


\section{\label{sec_application}Application Example}

\begin{figure}[!p]
\centering
\includegraphics{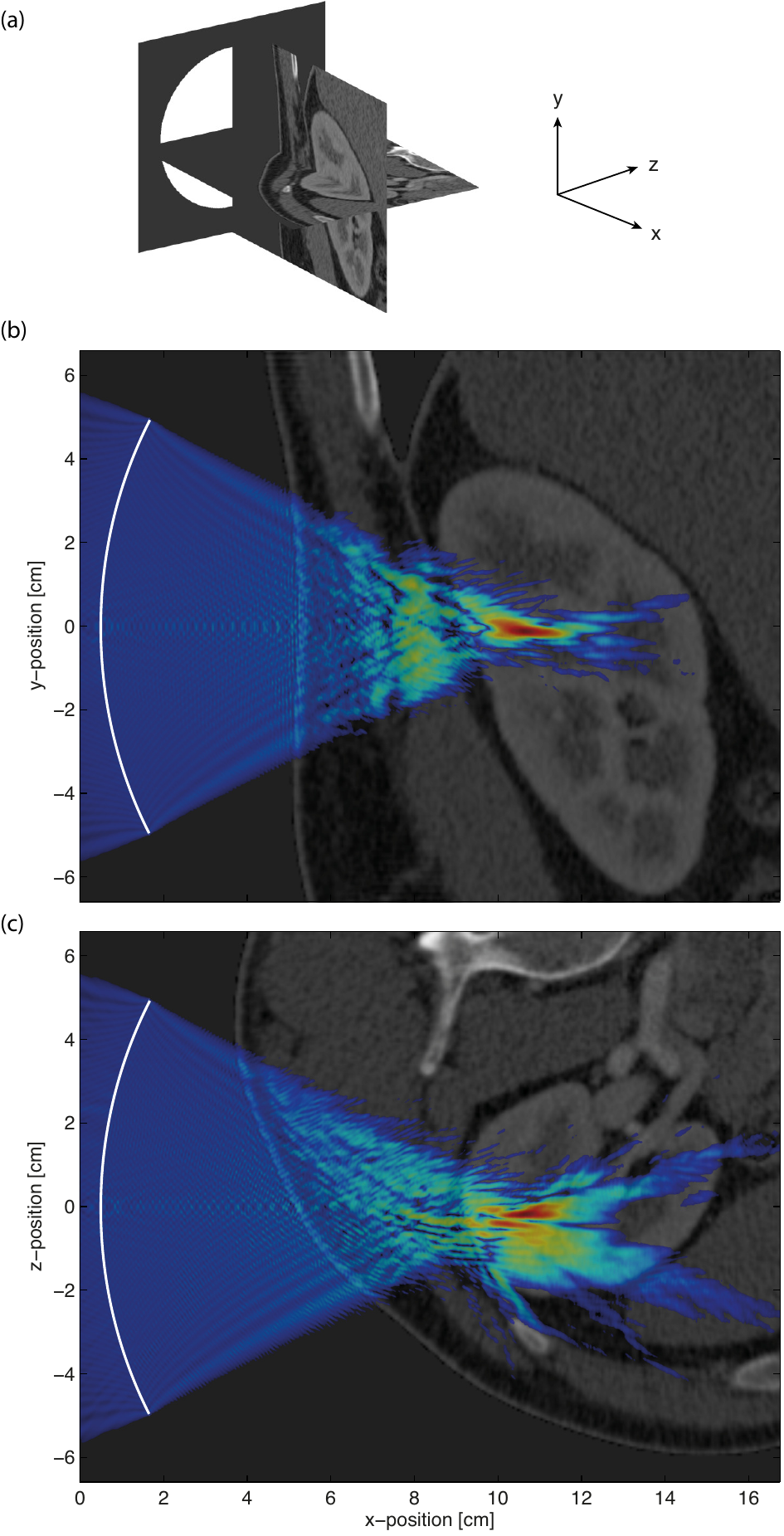}
\caption{(a) Sagittal and transverse slices through the abdominal CT scan used to define the material properties for simulating a HIFU treatment of the kidney. The approximate position of the HIFU transducer is shown with a white circle. (b)-(c) Saggittal and transverse slices through the simulated distribution of maximum pressure overlaid onto the corresponding CT slices. The pressure distribution is displayed using a log-scale and is thresholded at -30 dB. The distortion of the HIFU focus due to the body wall and the fat layer surrounding the kidney is clearly visible.}
\label{figure_kidney_hifu}
\end{figure}

To illustrate the utility of the developed nonlinear ultrasound model for solving real-world problems, a complete large-scale nonlinear ultrasound simulation representing a single HIFU sonication of the kidney was performed.  The medium properties for the simulation were derived from an abdominal CT scan (the MECANIX dataset available from \url{http://www.osirix-viewer.com/datasets/}). This was resampled using linear interpolation to give the appropriate resolution. The density of the tissue was calculated from Hounsfield units using the data from Schneider {\em et al.} \cite{Schneider1996}, and the sound speed was then estimated using the empirical relationship given by Mast \cite{Mast2000}. The remaining material properties were assigned book values \cite{Duck1990}. The HIFU transducer was defined as a circular bowl with a width of 10 cm and a focal length of 11 cm. The shape of the transducer within the 3D Cartesian grid was defined using a 3D extension of the midpoint circle algorithm \cite{Treeby2010}. The transducer was positioned behind the patient as shown in Fig.\ \ref{figure_kidney_hifu}(a), and was driven at 1 MHz by a continuous wave sinusoid. The acoustic intensity at the transducer surface was set to 2 W/cm$^2$ to simulate a treatment that would likely operate largely in a thermal regime (i.e., with minimal cavitation). Outside the body, the medium was assigned the properties of water \cite{Duck1990}. The total domain size was 17 cm $\times$ 14.3 cm $\times$ 14.3 cm and the grid spacing in each Cartesian direction was set to 93 $\mu$m, giving a total grid size of 2048 $\times$ 1536 $\times$ 1536 grid points and a maximum supported frequency of 8 MHz (i.e., eight harmonics of the source frequency). The simulation length was set to 220 $\mu s$ with a CFL number of 0.18, giving a total of 19800 time steps. Simulations of this scale and complexity have not previously been possible. 

The simulation was executed using 768 cores on VAYU with seven checkpoint-restart stages.  The total wall-clock time was 31 hours and 20 minutes, and the total memory used was 780 GB. The compressed input file was 45 GB, while the output file was 450 GB. This comprised 36 GB to store the peak positive and peak negative pressure across the domain, and 414 GB to store the time varying pressure and particle velocity for 6 periods in steady state over a 45 mm $\times$ 30 mm $\times$ 30 mm region surrounding the HIFU focus. Transverse and sagittal slices through the peak positive pressure overlaid onto the corresponding CT data used to define the material properties are shown in Fig.\ \ref{figure_kidney_hifu}(b)-(c). The distortion of the ultrasound focus due to the body wall and the fat layer surrounding the kidney is clearly visible. These effects have been noted clinically, and remain a barrier to the application of HIFU in the kidney \cite{Illing2005}. Thus, one possible future application of the developed $k$-space model would be a systematic investigation into the conditions necessary for viable HIFU ablation in the kidney (for example, the maximum thickness of the fat layer). In any case, the example serves to illustrate the utility of the implementation.


\section{Summary and Discussion}

A full-wave model for simulating the propagation of nonlinear ultrasound waves through absorbing and heterogeneous media is presented in the context of model-based treatment planning for HIFU. The governing equations are discretised using the $k$-space pseudospectral method. The model is implemented in C++ using MPI to enable large-scale problems to be solved using a distributed computer cluster. The performance of the model is evaluated using grid sizes up to 4096 $\times$ 2048 $\times$ 2048 grid points and up to 1024 compute cores. This is significantly larger than most ultrasound simulations previously presented, both in terms of grid size and the number of wavelengths this accurately represents. Given the global nature of the gradient calculation, the model shows good strong scaling behaviour, with a speed-up of 1.7x whenever the number of cores is doubled. This means large-scale problems can be spread over increasing numbers of compute cores with only a small computational overhead. The overall efficiency of the parallel implementation is on the order of 60\%, which corresponds to the ratio between computation and communication. The large communication overhead is due to the global all-to-all transposition that must be performed for every FFT (a second transposition is avoided by performing operations in the spatial frequency domain in transformed space). Finally, the efficacy of the model for studying real-world problems in HIFU is demonstrated using a large-scale simulation of a HIFU sonication of the kidney.

In the context of the large-scale problems outlined in Sec.\ 1 and Table \ref{TABLE_how_big}, the model developed here allows many problems of interest to be solved using a full-wave model for the first time. This is relevant for studying the aberration of HIFU beams in the body when the focal intensities are relatively low. This has many possible applications, for example, in treatment planning, exposimetry, patient selection, and equipment design. However, solving even larger problems involving high focal intensities where many 10's or 100's of harmonics may be present (as in Ref.\ \cite{Yuldashev2011}, for example) is still currently out of reach. Looking forward, there are two numerical strategies that might allow the model to be extended further. First, the governing equations could be solved on a non-uniform grid \cite{Treeby2013d}. In the current model, the grid spacing is globally constrained by the highest frequency harmonic that exists anywhere in the domain. However, in practice, the high-frequency harmonics are usually restricted to a small region near the ultrasound focus. Using a non-uniform grid would allow the grid points to be clustered around steep regions of the wavefield, and thus significantly reduce the total number of grid points needed for accurate simulations \cite{Treeby2013d}. Second, a domain decomposition approach could be used in which FFTs are computed locally on each partition using local Fourier basis \cite{Israeli1994,Albin2012}. This would replace the global communication required by the 3D FFT with local communications between processes responsible for neighbouring partitions. 

Considering the computer code, the main limitation is related to the 1D decomposition used to partition the domain. Although this approach exhibits relatively good scaling characteristics, it limits the maximum number of cooperating processes to be $P \leq \text{min} \left( \mathrm{N}_y, \mathrm{N}_z \right)$. This limitation is particularly relevant looking towards the exascale era, where supercomputers integrating over 1M cores are predicted to appear before 2020 \cite{Dongarra2011}. Being prepared for this sort of compute facility requires simulation tools that can efficiently employ hundreds of thousands of compute cores.  Moreover, while the trend in supercomputing is to integrate more and more compute cores, the total amount of memory is growing much more slowly \cite{Dongarra2011}. Effectively, this means the memory available per core will remain constant or even decrease in next generation systems. As an example, VAYU has 11936 cores with 3 GB/core, while it's successor RAIJIN has 57472 cores with only 2 GB/core. This is relevant because with the current 1D decomposition, the maximum grid size that can be solved  is ultimately limited by the memory per core. Both of these drawbacks could be solved by using a 2D partitioning approach where the 2D slabs are further broken into 1D pencils, with every process assigned a subset of pencils rather than a complete 2D slab. This would make higher numbers of compute cores (and thus memory) accessible to the simulation, where $P \leq \text{min} \left( \mathrm{N}_y^2, \mathrm{N}_z^2 \right)$. 

Another challenge for the current implementation is the amount of output data generated by the code.  When recording the time-varying acoustic pressure and particle velocity in a central region (e.g., near the HIFU focus), a single simulation can easily generate 0.5 TB of output data. Copying, post-processing, visualising, and archiving such large amounts of data quickly becomes impractical.  New techniques for on-the-fly post-processing are thus needed. The use of localised data sampling also introduces a work imbalance into the simulation code. If the output data is only collected from a small region of the domain, only a small subset of the processes actually store data to the disk, with the rest idle. The effective bandwidth to disk could thus be improved by redistributing the data to idle processes after it is collected, allowing more cores to be used for disk operations. Similarly, if some processes only collect a very small amount of data (e.g., from a single grid point in the local partition), the I/O subsystem can become congested by many small write requests resulting in poor performance. In this case, it would be better to collect the output data within each node before writing to disk. These improvements will be explored as part of future work.


\section*{Acknowledgments}
This work was supported by the Australian Research Council/Microsoft Linkage Project LP100100588. Computational resources were provided by the Australian National Computing Infrastructure (NCI) National Facility at the Australian National University (VAYU supercomputer), and iVEC at Murdoch University (EPIC supercomputer). The authors would like to thank Greg Clement for discussion on clinical applications of HIFU, and Ben Cox for helpful suggestions on the manuscript.


\section*{References}
\small
\bibliographystyle{unsrtnat-shortnames}
\bibliography{/Users/btreeby/Dropbox/BibTex/library}

\begin{thebibliography}{61}
\providecommand{\natexlab}[1]{#1}
\providecommand{\url}[1]{\texttt{#1}}
\expandafter\ifx\csname urlstyle\endcsname\relax
  \providecommand{\doi}[1]{doi: #1}\else
  \providecommand{\doi}{doi: \begingroup \urlstyle{rm}\Url}\fi

\bibitem[Kennedy et~al.(2003)Kennedy, ter Haar, and Cranston]{Kennedy2003}
J.~E. Kennedy, G.~R. ter Haar, and D.~Cranston.
\newblock {High intensity focused ultrasound: surgery of the future?}
\newblock \emph{Brit. J. Radiol.}, 76\penalty0 (909):\penalty0 590--599, 2003.

\bibitem[ter Haar(2007)]{TerHaar2007}
G.~ter Haar.
\newblock {Therapeutic applications of ultrasound}.
\newblock \emph{Prog. Biophys. Mol. Biol.}, 93\penalty0 (1-3):\penalty0
  111--129, 2007.

\bibitem[Clement(2004)]{Clement2004}
G.~T. Clement.
\newblock {Perspectives in clinical uses of high-intensity focused ultrasound}.
\newblock \emph{Ultrasonics}, 42\penalty0 (10):\penalty0 1087--93, 2004.

\bibitem[Jolesz and Hynynen(2008)]{Jolesz2008}
F.~A. Jolesz and K.~H. Hynynen.
\newblock \emph{{MRI-guided focused ultrasound surgery}}.
\newblock Informa Healthcare, New York, 2008.

\bibitem[Zhang and Wang(2010)]{Zhang2010a}
L.~Zhang and Z.-B. Wang.
\newblock {High-intensity focused ultrasound tumor ablation: review of ten
  years of clinical experience}.
\newblock \emph{Front. Med. China}, 4\penalty0 (3):\penalty0 294--302, 2010.

\bibitem[Paulides et~al.(2013)Paulides, Stauffer, Neufeld, Maccarini, Kyriakou,
  Canters, Diederich, Bakker, and {Van Rhoon}]{Paulides2013}
M.~M. Paulides, P.~R. Stauffer, E.~Neufeld, P.~F. Maccarini, A.~Kyriakou,
  R.~A.~M. Canters, C.~J. Diederich, J.~F. Bakker, and G.~C. {Van Rhoon}.
\newblock {Simulation techniques in hyperthermia treatment planning}.
\newblock \emph{Int. J. Hyperthermia}, 29\penalty0 (4):\penalty0 346--357,
  2013.

\bibitem[Liu et~al.(2005)Liu, McDannold, and Hynynen]{Liu2005}
H.-L. Liu, N.~McDannold, and K.~Hynynen.
\newblock {Focal beam distortion and treatment planning in abdominal focused
  ultrasound surgery}.
\newblock \emph{Med. Phys.}, 32\penalty0 (5):\penalty0 1270--1280, 2005.

\bibitem[Connor and Hynynen(2002)]{Connor2002}
C.~W. Connor and K.~Hynynen.
\newblock {Bio-acoustic thermal lensing and nonlinear propagation in focused
  ultrasound surgery using large focal spots: a parametric study.}
\newblock \emph{Phys. Med. Biol.}, 47\penalty0 (11):\penalty0 1911--28, 2002.

\bibitem[Lepock(2003)]{Lepock2003}
J.~R. Lepock.
\newblock {Cellular effects of hyperthermia: relevance to the minimum dose for
  thermal damage.}
\newblock \emph{Int. J. Hyperthermia}, 19\penalty0 (3):\penalty0 252--66, 2003.

\bibitem[Wojcik et~al.(1995)Wojcik, Mould, Abboud, Ostromogilsky, and
  Vaughan]{Wojcik1995}
G.~L. Wojcik, J.~Mould, N.~Abboud, M.~Ostromogilsky, and D.~Vaughan.
\newblock {Nonlinear modeling of therapeutic ultrasound}.
\newblock In \emph{IEEE International Ultrasonics Symposium}, pages 1617--1622,
  1995.

\bibitem[Khokhlova et~al.(2010)Khokhlova, Bessonova, Soneson, Canney, Bailey,
  and Crum]{Khokhlova2010}
V.~A. Khokhlova, O.~V. Bessonova, J.~E. Soneson, M.~S. Canney, M.~R. Bailey,
  and L.~A. Crum.
\newblock {Bandwidth limitations in characterization of high intensity focused
  ultrasound fields in the presence of shocks}.
\newblock In \emph{9th International Symposium on Therapeutic Ultrasound},
  pages 363--366, 2010.

\bibitem[Averkiou and Cleveland(1999)]{Averkiou1999}
M.~A. Averkiou and R.~O. Cleveland.
\newblock {Modeling of an electrohydraulic lithotripter with the KZK equation.}
\newblock \emph{J. Acoust. Soc. Am.}, 106\penalty0 (1):\penalty0 102--112,
  1999.

\bibitem[Curra et~al.(2000)Curra, Mourad, Khokhlova, Cleveland, and
  Crum]{Curra2000}
F.~P. Curra, P.~D. Mourad, V.~A. Khokhlova, R.~O. Cleveland, and L.~A. Crum.
\newblock {Numerical simulations of heating patterns and tissue temperature
  response due to high-intensity focused ultrasound}.
\newblock \emph{IEEE Trans. Ultrason. Ferroelectr. Freq. Control}, 47\penalty0
  (4):\penalty0 1077--89, 2000.

\bibitem[Khokhlova et~al.(2001)Khokhlova, Souchon, Tavakkoli, Sapozhnikov, and
  Cathignol]{Khokhlova2001}
V.~A. Khokhlova, R.~Souchon, J.~Tavakkoli, O.~A. Sapozhnikov, and D.~Cathignol.
\newblock {Numerical modeling of finite-amplitude sound beams: Shock formation
  in the near field of a cw plane piston source}.
\newblock \emph{J. Acoust. Soc. Am.}, 110\penalty0 (1):\penalty0 95--108, 2001.

\bibitem[Yuldashev and Khokhlova(2011)]{Yuldashev2011}
P.~V. Yuldashev and V.~A. Khokhlova.
\newblock {Simulation of three-dimensional nonlinear fields of ultrasound
  therapeutic arrays}.
\newblock \emph{Acoust. Phys.}, 57\penalty0 (3):\penalty0 334--343, 2011.

\bibitem[Yuldashev et~al.(2010)Yuldashev, Krutyansky, Khokhlova, Brysev, and
  Bunkin]{Yuldashev2010}
P.~V. Yuldashev, L.~M. Krutyansky, V.~A. Khokhlova, A.~P. Brysev, and F.~V.
  Bunkin.
\newblock {Distortion of the focused finite amplitude ultrasound beam behind
  the random phase layer}.
\newblock \emph{Acoust. Phys.}, 56\penalty0 (4):\penalty0 467--474, 2010.

\bibitem[Pinton et~al.(2009)Pinton, Dahl, Rosenzweig, and Trahey]{Pinton2009}
G.~F. Pinton, J.~Dahl, S.~Rosenzweig, and G.~E. Trahey.
\newblock {A heterogeneous nonlinear attenuating full-wave model of
  ultrasound}.
\newblock \emph{IEEE Trans. Ultrason. Ferroelectr. Freq. Control}, 56\penalty0
  (3):\penalty0 474--488, 2009.

\bibitem[Okita et~al.(2011)Okita, Ono, Takagi, and Matsumoto]{Okita2011}
K.~Okita, K.~Ono, S.~Takagi, and Y.~Matsumoto.
\newblock {Development of high intensity focused ultrasound simulator for
  large-scale computing}.
\newblock \emph{Int. J. Numer. Meth. Fluids}, 65:\penalty0 43--66, 2011.

\bibitem[Pinton et~al.(2011)Pinton, Aubry, Fink, and Tanter]{Pinton2011a}
G.~Pinton, J.-F. Aubry, M.~Fink, and M.~Tanter.
\newblock {Effects of nonlinear ultrasound propagation on high intensity brain
  therapy}.
\newblock \emph{Med. Phys.}, 38\penalty0 (3):\penalty0 1207--1216, 2011.

\bibitem[Okita et~al.(2014)Okita, Narumi, Azuma, Takagi, and
  Matumoto]{Okita2014}
K.~Okita, R.~Narumi, T.~Azuma, S.~Takagi, and Y.~Matumoto.
\newblock {The role of numerical simulation for the development of an advanced
  HIFU system}.
\newblock \emph{Comput. Mech.}, 2014.

\bibitem[Pulkkinen et~al.(2014)Pulkkinen, Werner, Martin, and
  Hynynen]{Pulkkinen2014a}
A.~Pulkkinen, B.~Werner, E.~Martin, and K.~Hynynen.
\newblock {Numerical simulations of clinical focused ultrasound functional
  neurosurgery}.
\newblock \emph{Phys. Med. Biol.}, 59\penalty0 (7):\penalty0 1679--1700, 2014.

\bibitem[Hesthaven et~al.(2007)Hesthaven, Gottlieb, and
  Gottlieb]{Hesthaven2007}
J.~S. Hesthaven, S.~Gottlieb, and D.~Gottlieb.
\newblock \emph{{Spectral Methods for Time-Dependent Problems}}.
\newblock Cambridge University Press, Cambridge, 2007.

\bibitem[Mast et~al.(2001)Mast, Souriau, Liu, Tabei, Nachman, and
  Waag]{Mast2001}
T.~D. Mast, L.~P. Souriau, D.-L.~D. Liu, M.~Tabei, A.~I. Nachman, and R.~C.
  Waag.
\newblock {A k-space method for large-scale models of wave propagation in
  tissue}.
\newblock \emph{IEEE Trans. Ultrason. Ferroelectr. Freq. Control}, 48\penalty0
  (2):\penalty0 341--354, 2001.

\bibitem[Tabei et~al.(2002)Tabei, Mast, and Waag]{Tabei2002}
M.~Tabei, T.~D. Mast, and R.~C. Waag.
\newblock {A k-space method for coupled first-order acoustic propagation
  equations}.
\newblock \emph{J. Acoust. Soc. Am.}, 111\penalty0 (1):\penalty0 53--63, 2002.

\bibitem[Haber et~al.(1973)Haber, Lee, Klein, and Boris]{Haber1973}
I.~Haber, R.~Lee, H.~H. Klein, and J.~P. Boris.
\newblock {Advances in electromagnetic plasma simulation techniques}.
\newblock In \emph{Proc. Sixth Conf. Num. Sim. Plasmas}, pages 46--48, 1973.

\bibitem[Bojarski(1982)]{Bojarski1982}
N.~N. Bojarski.
\newblock {The k-space formulation of the scattering problem in the time
  domain}.
\newblock \emph{J. Acoust. Soc. Am.}, 72\penalty0 (2):\penalty0 570--584, 1982.

\bibitem[Bojarski(1985)]{Bojarski1985}
N.~N. Bojarski.
\newblock {The k-space formulation of the scattering problem in the time
  domain: An improved single propagator formulation}.
\newblock \emph{J. Acoust. Soc. Am.}, 77\penalty0 (3):\penalty0 826--831, 1985.

\bibitem[Fornberg(1987)]{Fornberg1987}
B.~Fornberg.
\newblock {The pseudospectral method: Comparisons with finite differences for
  the elastic wave equation}.
\newblock \emph{Geophysics}, 52\penalty0 (4):\penalty0 483--501, 1987.

\bibitem[Liu(1999)]{Liu1999b}
Q.~H. Liu.
\newblock {Large-scale simulations of electromagnetic and acoustic measurements
  using the pseudospectral time-domain (PSTD) algorithm}.
\newblock \emph{IEEE. T. Geosci. Remote}, 37\penalty0 (2):\penalty0 917--926,
  1999.

\bibitem[Cox et~al.(2007)Cox, Kara, Arridge, and Beard]{Cox2007}
B.~T. Cox, S.~Kara, S.~R. Arridge, and P.~C. Beard.
\newblock {k-space propagation models for acoustically heterogeneous media:
  Application to biomedical photoacoustics}.
\newblock \emph{J. Acoust. Soc. Am.}, 121\penalty0 (6):\penalty0 3453--3464,
  2007.

\bibitem[Daoud and Lacefield(2009)]{Daoud2009}
M.~I. Daoud and J.~C. Lacefield.
\newblock {Distributed three-dimensional simulation of B-mode ultrasound
  imaging using a first-order k-space method}.
\newblock \emph{Phys. Med. Biol.}, 54\penalty0 (17):\penalty0 5173--5192, 2009.

\bibitem[Tillett et~al.(2009)Tillett, Daoud, Lacefield, and Waag]{Tillett2009}
J.~C. Tillett, M.~I. Daoud, J.~C. Lacefield, and R.~C. Waag.
\newblock {A k-space method for acoustic propagation using coupled first-order
  equations in three dimensions}.
\newblock \emph{J. Acoust. Soc. Am.}, 126\penalty0 (3):\penalty0 1231--1244,
  2009.

\bibitem[Wojcik et~al.(1998)Wojcik, Mould, Ayter, and Carcione]{Wojcik1998}
G.~Wojcik, J.~Mould, S.~Ayter, and L.~Carcione.
\newblock {A study of second harmonic generation by focused medical transducer
  pulses}.
\newblock In \emph{IEEE International Ultrasonics Symposium}, pages 1583--1588,
  1998.

\bibitem[Treeby et~al.(2011)Treeby, Tumen, and Cox]{Treeby2011a}
B.~E. Treeby, M.~Tumen, and B.~T. Cox.
\newblock {Time domain simulation of harmonic ultrasound images and beam
  patterns in 3D using the k-space pseudospectral method}.
\newblock In \emph{Medical Image Computing and Computer-Assisted Intervention,
  Part I}, volume 6891, pages 363--370. Springer, Heidelberg, 2011.

\bibitem[Albin et~al.(2012)Albin, Bruno, Cheung, and Cleveland]{Albin2012}
N.~Albin, O.~P. Bruno, T.~Y. Cheung, and R.~O. Cleveland.
\newblock {Fourier continuation methods for high-fidelity simulation of
  nonlinear acoustic beams.}
\newblock \emph{J. Acoust. Soc. Am.}, 132\penalty0 (4):\penalty0 2371--87,
  2012.

\bibitem[Jing et~al.(2012)Jing, Wang, and Clement]{Jing2012b}
Y.~Jing, T.~Wang, and G.~T. Clement.
\newblock {A k-space method for moderately nonlinear wave propagation.}
\newblock \emph{IEEE Trans. Ultrason. Ferroelectr. Freq. Control}, 59\penalty0
  (8):\penalty0 1664--1673, 2012.

\bibitem[Treeby et~al.(2012)Treeby, Jaros, Rendell, and Cox]{Treeby2012}
B.~E. Treeby, J.~Jaros, A.~P. Rendell, and B.~T. Cox.
\newblock {Modeling nonlinear ultrasound propagation in heterogeneous media
  with power law absorption using a k-space pseudospectral method}.
\newblock \emph{J. Acoust. Soc. Am.}, 131\penalty0 (6):\penalty0 4324--4336,
  2012.

\bibitem[Verweij and Huijssen(2009)]{Verweij2009}
M.~D. Verweij and J.~Huijssen.
\newblock {A filtered convolution method for the computation of acoustic wave
  fields in very large spatiotemporal domains}.
\newblock \emph{J. Acoust. Soc. Am.}, 125\penalty0 (4):\penalty0 1868--1878,
  2009.

\bibitem[Huijssen and Verweij(2010)]{Huijssen2010}
J.~Huijssen and M.~D. Verweij.
\newblock {An iterative method for the computation of nonlinear, wide-angle,
  pulsed acoustic fields of medical diagnostic transducers}.
\newblock \emph{J. Acoust. Soc. Am.}, 127\penalty0 (1):\penalty0 33--44, 2010.

\bibitem[Demi et~al.(2011)Demi, van Dongen, and Verweij]{Demi2011}
L.~Demi, K.~W.~A. van Dongen, and M.~D. Verweij.
\newblock {A contrast source method for nonlinear acoustic wave fields in media
  with spatially inhomogeneous attenuation}.
\newblock \emph{J. Acoust. Soc. Am.}, 129\penalty0 (3):\penalty0 1221--1230,
  2011.

\bibitem[Westervelt(1963)]{Westervelt1963}
P.~J. Westervelt.
\newblock {Parametric acoustic array}.
\newblock \emph{J. Acoust. Soc. Am.}, 35\penalty0 (4):\penalty0 535--537, 1963.

\bibitem[Hamilton and Morfey(2008)]{Hamilton2008}
M.~F. Hamilton and C.~L. Morfey.
\newblock {Model Equations}.
\newblock In M.~F. Hamilton and D.~T. Blackstock, editors, \emph{Nonlinear
  Acoustics}, pages 41--63. Acoustical Society of America, New York, 2008.

\bibitem[Aanonsen et~al.(1984)Aanonsen, Barkve, Tjotta, and
  Tjotta]{Aanonsen1984}
S.~I. Aanonsen, T.~Barkve, J.~N. Tjotta, and S.~Tjotta.
\newblock {Distortion and harmonic generation in the nearfield of a finite
  amplitude sound beam}.
\newblock \emph{J. Acoust. Soc. Am.}, 75\penalty0 (3):\penalty0 749--768, 1984.

\bibitem[Duck(1990)]{Duck1990}
F.~A. Duck.
\newblock \emph{{Physical Properties of Tissue: A Comprehensive Reference
  Book}}.
\newblock Academic Press, 1990.

\bibitem[Chen and Holm(2004)]{Chen2004}
W.~Chen and S.~Holm.
\newblock {Fractional Laplacian time-space models for linear and nonlinear
  lossy media exhibiting arbitrary frequency power-law dependency}.
\newblock \emph{J. Acoust. Soc. Am.}, 115\penalty0 (4):\penalty0 1424--1430,
  2004.

\bibitem[Treeby and Cox(2010{\natexlab{a}})]{Treeby2010a}
B.~E. Treeby and B.~T. Cox.
\newblock {Modeling power law absorption and dispersion for acoustic
  propagation using the fractional Laplacian}.
\newblock \emph{J. Acoust. Soc. Am.}, 127\penalty0 (5):\penalty0 2741--2748,
  2010{\natexlab{a}}.

\bibitem[Treeby and Cox(2011)]{Treeby2011d}
B.~E. Treeby and B.~T. Cox.
\newblock {A k-space Green’s function solution for acoustic initial value
  problems in homogeneous media with power law absorption}.
\newblock \emph{J. Acoust. Soc. Am.}, 129\penalty0 (6):\penalty0 3652--3660,
  2011.

\bibitem[Berenger(1996)]{Berenger1996}
J.-P. Berenger.
\newblock {Three-dimensional perfectly matched layer for the absorption of
  electromagnetic waves}.
\newblock \emph{J. Comput. Phys.}, 127\penalty0 (2):\penalty0 363--379, 1996.

\bibitem[Frigo and Johnson(2005)]{Frigo2005}
M.~Frigo and S.~G. Johnson.
\newblock {The design and implementation of FFTW3}.
\newblock \emph{Proc. IEEE}, 93\penalty0 (2):\penalty0 216--231, 2005.

\bibitem[Pekurovsky(2012)]{Pekurovsky2012}
D.~Pekurovsky.
\newblock {P3DFFT: A framework for parallel computations of Fourier transforms
  in three dimensions}.
\newblock \emph{SIAM J. Sci. Comput.}, 34\penalty0 (4):\penalty0 C192--C209,
  2012.

\bibitem[Yeung et~al.(2012)Yeung, Donzis, and Sreenivasan]{Yeung2012}
P.~K. Yeung, D.~A. Donzis, and K.~R. Sreenivasan.
\newblock {Dissipation, enstrophy and pressure statistics in turbulence
  simulations at high Reynolds numbers}.
\newblock \emph{Journal of Fluid Mechanics}, 700:\penalty0 5--15, 2012.

\bibitem[Schneider et~al.(1996)Schneider, Pedroni, and Lomax]{Schneider1996}
U.~Schneider, E.~Pedroni, and A.~Lomax.
\newblock {The calibration of CT Hounsfield units for radiotherapy treatment
  planning.}
\newblock \emph{Phys. Med. Biol.}, 41\penalty0 (1):\penalty0 111--124, 1996.

\bibitem[ter Haar et~al.(2011)ter Haar, Shaw, Pye, Ward, Bottomley, Nolan, and
  Coady]{TerHaar2011}
G.~ter Haar, A.~Shaw, S.~Pye, B.~Ward, F.~Bottomley, R.~Nolan, and A.-M. Coady.
\newblock {Guidance on reporting ultrasound exposure conditions for bio-effects
  studies}.
\newblock \emph{Ultrasound Med. Biol.}, 37\penalty0 (2):\penalty0 177--183,
  2011.

\bibitem[Treeby and Cox(2010{\natexlab{b}})]{Treeby2010}
B.~E. Treeby and B.~T. Cox.
\newblock {k-Wave: MATLAB toolbox for the simulation and reconstruction of
  photoacoustic wave fields}.
\newblock \emph{J. Biomed. Opt.}, 15\penalty0 (2):\penalty0 021314,
  2010{\natexlab{b}}.

\bibitem[Frigo and Johnson(2012)]{Frigo2012a}
M.~Frigo and S.~G. Johnson.
\newblock {FFTW user manual}.
\newblock Technical Report November, Massachusetts Institute of Technology,
  Boston, 2012.

\bibitem[Jaros et~al.(2012)Jaros, Treeby, and Rendell]{Jaros2012}
J.~Jaros, B.~E. Treeby, and A.~P. Rendell.
\newblock {Use of multiple GPUs on shared memory multiprocessors for ultrasound
  propagation simulations}.
\newblock In J.~Chen and R.~Ranjan, editors, \emph{10th Australasian Symposium
  on Parallel and Distributed Computing}, volume 127, pages 43--52. ACS, 2012.

\bibitem[Mast(2000)]{Mast2000}
T.~D. Mast.
\newblock {Empirical relationships between acoustic parameters in human soft
  tissues}.
\newblock \emph{Acoustics Research Letters Online}, 1\penalty0 (2):\penalty0
  37--42, 2000.

\bibitem[Illing et~al.(2005)Illing, Kennedy, Wu, ter Haar, Protheroe, Friend,
  Gleeson, Cranston, Phillips, and Middleton]{Illing2005}
R.~O. Illing, J.~E. Kennedy, F.~Wu, G.~R. ter Haar, A.~S. Protheroe, P.~J.
  Friend, F.~V. Gleeson, D.~W. Cranston, R.~R. Phillips, and M.~R. Middleton.
\newblock {The safety and feasibility of extracorporeal high-intensity focused
  ultrasound (HIFU) for the treatment of liver and kidney tumours in a Western
  population.}
\newblock \emph{Brit. J. Cancer}, 93\penalty0 (8):\penalty0 890--895, 2005.

\bibitem[Treeby(2013)]{Treeby2013d}
B.~E. Treeby.
\newblock {Modeling nonlinear wave propagation on nonuniform grids using a
  mapped k-space pseudospectral method.}
\newblock \emph{IEEE Trans. Ultrason. Ferroelectr. Freq. Control}, 60\penalty0
  (10):\penalty0 2208--2013, 2013.

\bibitem[Israeli et~al.(1994)Israeli, Vozovoi, and Averbuch]{Israeli1994}
M.~Israeli, L.~Vozovoi, and A.~Averbuch.
\newblock {Domain decomposition methods with local fourier basis for parabolic
  problems}.
\newblock \emph{Contemp. Math}, 157:\penalty0 223--230, 1994.

\bibitem[Dongarra et~al.(2011)Dongarra, Beckman, Moore, Aerts, Aloisio, Andre,
  Barkai, Berthou, Boku, Braunschweig, Cappello, Chapman, Choudhary, Dosanjh,
  Dunning, Fiore, Geist, Gropp, Harrison, Hereld, Heroux, Hoisie, Hotta,
  Ishikawa, Johnson, Kale, Kenway, Keyes, Kramer, Labarta, Lichnewsky, Lippert,
  Lucas, Maccabe, Matsuoka, Messina, Michielse, Mohr, Mueller, Nagel,
  Nakashima, Papka, Reed, Sato, Seidel, Shalf, Skinner, Snir, Sterling,
  Stevens, Streitz, Sugar, Sumimoto, Tang, Taylor, Thakur, Trefethen, Valero,
  van~der Steen, Vetter, Williams, Wisniewski, and Yelick]{Dongarra2011}
J.~Dongarra, P.~Beckman, T.~Moore, P.~Aerts, G.~Aloisio, J.-C. Andre,
  D.~Barkai, J.-Y. Berthou, T.~Boku, B.~Braunschweig, F.~Cappello, B.~Chapman,
  A.~Choudhary, S.~Dosanjh, T.~Dunning, S.~Fiore, A.~Geist, B.~Gropp,
  R.~Harrison, M.~Hereld, M.~Heroux, A.~Hoisie, K.~Hotta, Y.~Ishikawa,
  F.~Johnson, S.~Kale, R.~Kenway, D.~Keyes, B.~Kramer, J.~Labarta,
  A.~Lichnewsky, T.~Lippert, B.~Lucas, B.~Maccabe, S.~Matsuoka, P.~Messina,
  P.~Michielse, B.~Mohr, M.~S. Mueller, W.~E. Nagel, H.~Nakashima, M.~E. Papka,
  D.~Reed, M.~Sato, E.~Seidel, J.~Shalf, D.~Skinner, M.~Snir, T.~Sterling,
  R.~Stevens, F.~Streitz, B.~Sugar, S.~Sumimoto, W.~Tang, J.~Taylor, R.~Thakur,
  A.~Trefethen, M.~Valero, A.~van~der Steen, J.~Vetter, P.~Williams,
  R.~Wisniewski, and K.~Yelick.
\newblock {The International Exascale Software Project roadmap}.
\newblock \emph{Int. J. High Perform. Comput. Appl.}, 25\penalty0 (1):\penalty0
  3--60, 2011.

\end{thebibliography}

\end{document}